%% file: ordinal.tex
\documentclass{article}

\author{Michael J. Wurm, Paul J. Rathouz, and Bret M. Hanlon\\\\University of Wisconsin--Madison}
\title{Regularized Ordinal Regression and the \textbf{ordinalNet} \textsf{R} Package}

\input{header}  

\begin{document}
	\maketitle
	
	\begin{abstract}
		Regularization techniques such as the lasso \citep{Tibshirani1996} and elastic net \citep{Zou2005} can be used to improve regression model coefficient estimation and prediction accuracy, as well as to perform variable selection. Ordinal regression models are widely used in applications where the use of regularization could be beneficial;  however, these models are not included in many popular software packages for regularized regression. We propose a coordinate descent algorithm to fit a broad class of ordinal regression models with an elastic net penalty. Furthermore, we demonstrate that each model in this class generalizes to a more flexible form, for instance to accommodate unordered categorical data. We introduce an elastic net penalty class that applies to both model forms. Additionally, this penalty can  be used to shrink a non-ordinal model toward its ordinal counterpart. Finally, we introduce the \textsf{R} package \textbf{ordinalNet}, which implements the algorithm for this model class.
	\end{abstract}


	\input{ordinal-text}
\end{document}

%% file: header.tex
\newcommand{\commentout}[1]{}

\newcommand{\ps}{parametrizations }

\usepackage{relsize}
\newcommand\CC{C\nolinebreak[4]\hspace{-.01em}\raisebox{.4ex}{\relsize{-3}{\textbf{++}}} }

\usepackage{natbib}
\usepackage{mathtools, amsmath, amsfonts, amssymb, bbm, bm}
\usepackage{graphicx}
\usepackage{rotating}
\usepackage{float}  
\usepackage{verbatim}  
\usepackage{hyperref}  
\usepackage{subfigure, color, latexsym, ifthen,commath,setspace}

\newboolean{includefigs}
\setboolean{includefigs}{true} 
\newcommand{\condcomment}[2]{\ifthenelse{#1}{#2}{}}

\newcommand{\beq}{\begin{equation}}
\newcommand{\eeq}{\end{equation}}

\newcommand{\be}{\begin{eqnarray*}}
\newcommand{\ee}{\end{eqnarray*}}

\def\tran{\mathop{ T }}



%% file: ordinal-text.tex

\section{Introduction}\label{sec:intro}
Ordinal regression models arise in contexts where the response variable belongs to one of several ordered categories (such as 1=``poor'', 2=``fair'', 3=``good'', 4=``excellent''). One of the most common regression models for this type of data is the cumulative logit model \citep{McCullagh1980}, which is also known as the proportional odds model or the ordinal logistic regression model. Other ordinal regression models include the stopping ratio model, the continuation ratio model, and the adjacent category model. The \textbf{VGAM} \textsf{R} package \citep{Yee1996, Yee2010, Yee2015} fits all of the aforementioned models, but without regularization or variable selection. The \textsf{SAS} CATMOD procedure also fits some of these models \citep{sas}. Popular CRAN packages for penalized regression, such as \textbf{penalized} \citep{goeman2014} and \textbf{glmnet} \citep{Friedman2010}, do not currently fit ordinal models.

Some algorithms and software already exist for penalized ordinal regression models. The \textsf{R} package \textbf{lrm} \citep{harrell2015} fits the cumulative logit model with quadratic (ridge regression) penalty. The \textsf{R} packages \textbf{glmnetcr} \citep{Archer2014a} and \textbf{glmpathcr} \citep{Archer2014b} fit stopping ratio models with the elastic net penalty. \citet{Archer2014a, Archer2014b} refers to these as continuation ratio models, but we define stopping ratio and continuation ratio models in the same way as \citet{Yee2010}.

\citet{Archer2014} also implemented the generalized monotone incremental forward stagewise (GMIFS) algorithm for regularized ordinal regression models in the \textsf{R} package \textbf{ordinalgmifs}. This procedure finds a solution path similar to the $L_1$ norm (lasso) penalty. In fact it is the same solution path if the lasso path is monotone for each coefficient, but in other cases the GMIFS and lasso solution paths differ \citep{Hastie2009}. Some drawbacks of this algorithm are that it fits a single solution path and does not have the flexibility of the elastic net mixing parameter (usually denoted by $\alpha$). It can also be computationally expensive because the entire solution path must be fit in small increments, whereas the lasso and elastic net solution path can be obtained only at specified values of the regularization tuning parameter (usually denoted by $\lambda$). A sequence of, say, twenty values may be enough to tune a model by cross validation and will usually be faster than fitting a longer sequence.

To summarize, algorithms for ordinal regression either do not allow regularization, or they apply to specific models. Hence, options are limited for ordinal regression with a large number of predictors. In that context, our contribution to this growing body of software and literature is threefold. First, we propose a general coordinate descent algorithm to fit a rich class of multinomial regression models, both ordinal and non-ordinal, with elastic net penalty.

Second, we define a class of models that (a) can be fit with the elastic net penalty by the aforementioned algorithm, (b) contains some of the most common ordinal regression models, (c) is convenient for modularizing the fitting algorithm, and (d) has both a parallel (ordinal) and a nonparallel form for each model (discussed in the next paragraph). We call this the \emph{elementwise link multinomial-ordinal} (ELMO) class of models. This class is a subset of vector generalized linear models \citep{Yee2015}. Each model in this class uses a multivariate link function to link multinomial probabilities to a set of linear predictors. The link function can be conveniently written as a composite of two functions. The first determines the model family (e.g. cumulative probability, stopping ratio, continuation ratio, or adjacent category). The second is a standard link function (e.g. logit, probit, or complementary log-log), which is applied elementwise to the result of the first function.

 Another feature of the ELMO class is that each model has a form that is appropriate for ordinal response data, as well as a more flexible form that can be applied to either ordinal or unordered categorical responses. We will refer to the first as the \emph{parallel} form and the second as the \emph{nonparallel} form. For the parallel form, the linear predictors of a given observation only differ by their intercept values---the other coefficients are the same. This restriction is what \citet{Yee2010} refers to as the parallelism assumption. The nonparallel form allows all of the coefficients to vary. An example from the ELMO class is the proportional odds model, which is a parallel model that has a nonparallel counterpart, the partial proportional odds model \citep{Peterson1990}. For more details, see Section \ref{subsec:forms}.

Finally, we propose an elastic net penalty class that applies to both the parallel and nonparallel forms. It can also be used to shrink the nonparallel model toward its parallel counterpart. This can be useful in a situation where one would like to fit an ordinal model but relax the parallelism assumption. This can be achieved by over-parameterizing the nonparallel model to include both the nonparallel and parallel coefficients. We call this alternate parametrization the \emph{semi-parallel} model. Although the regression model itself is not identifiable under this parametrization, the penalized likelihood has a unique optimum (or almost unique, as discussed in Appendix \ref{app:uniqueness}).

We provide an outline for the remainder of the work. Section \ref{sec:elmo} defines the ELMO class with specific examples. We also define the parallel, nonparallel, and semi-parallel \ps with the elastic net penalty. Section \ref{sec:cd} provides the proposed algorithm for fitting multinomial regression models with the elastic net penalty. Section \ref{sec:sim} presents a simulation study to compare prediction accuracy of the penalized parallel, nonparallel, and semi-parallel models. Section \ref{sec:compare} demonstrates the use of penalized ELMO class models for out-of-sample prediction and variable selection alongside other methods. Section \ref{sec:package} provides details about the \textbf{ordinalNet} \textsf{R} package, which is available on the Comprehensive \textsf{R} Archive Network. Section \ref{sec:demo} provides a demonstration of the \textbf{ordinalNet} package. Section \ref{sec:discussion} contains a summary of the findings and contribution.

\section{Elementwise link multinomial-ordinal (ELMO) class}\label{sec:elmo}
This section is organized as follows. Section \ref{subsec:notation} introduces commonly used notation. Section \ref{subsec:intro-elmo} is the heart of Section \ref{sec:elmo}, defining  the ELMO model class. The remaining subsections then discuss particular elements of the ELMO class and issues related to elastic net penalization of this class. Sections \ref{subsec:family} and \ref{subsec:element} provide details for the family function and elementwise link function, respectively. The parallel, nonparallel, and semi-parallel forms are discussed in Sections \ref{subsec:forms} and \ref{subsec:semi}. Section \ref{subsec:ENP} discusses the elastic net penalty and formulates the objective function under the three model forms. Finally, Section \ref{subsec:ILFJ} describes the Jacobian of the inverse link function for the ELMO class.

\subsection{Notation}\label{subsec:notation}
We introduce commonly used notation. Paper-specific notation is developed throughout the work. For a vector $x$, we use $x^{\tran}$ to denote its transpose. For a matrix $A$ we use bracket notation to indicate elements, so that $\left[ A \right]_{ij}$, indicates the $(i,j)^{th}$ element of $A$.  $\mathbbm{1}_K = \mathbbm{1}$ denotes  the length-$K$ column vector of ones and $I_{K \times K} = I$ denotes the $K \times K$ identity matrix. In both cases, the dependence on $K$ will be suppressed when it is clear from the context. We use $\nabla$ for the gradient operator and $D$ for the Jacobian operator. Consider a vector-valued function $f$ with vector argument $x$. As is standard, the Jacobian of $f$ is defined as
$$
D f(x) = \frac{\partial f(x)}{\partial x^{\tran}},
$$
in other words
$$
[D f(x)]_{mn} = \frac{\partial f(x)_m }{\partial x_n}.
$$

\subsection{An Introduction to the ELMO model class}\label{subsec:intro-elmo}
We now define the ELMO model class. Models within this class are completely specified by their multivariate link function, which is a composite of two functions. The first function determines the model family (e.g. cumulative probability, stopping ratio, continuation ratio, or adjacent category). We refer to these as \emph{multinomial-ordinal} (MO) families because each has a parallel form specifically for ordinal data, as well as a nonparallel form for any multinomial data, ordinal or unordered. The second function is an \emph{elementwise link} (EL) function, which applies a standard link function on $(0, 1) \rightarrow \mathbb{R}$ (e.g. logit, probit, or complementary log-log) elementwise to the result of the first function. \citet{McCullagh1980}, \citet{Wilhelm1998}, and \citet{Yee2010} provide more background on categorical regression models.

Let $\bm{y}_i|\bm{x}_i=x_i \overset{\text{indep.}}{\sim} \text{Multinomial}(n_i, p_{i1}, p_{i2}, \ldots, p_{i(K+1)})$ for $i=1,\dots,N$. Observations (e.g. subjects or patients) are indexed by $i$, and $N$ is the total number of observations. Here, $x_i$ is an observed vector of covariates (without an intercept), and $\bm{y}_i = (\bm{y}_{i1}, \ldots, \bm{y}_{i(K+1)})^T$ is a random vector of counts summing to $n_i$. The conditional distribution represents $n_i$ independent trials which fall into $K+1$ classes with probabilities $(p_{i1}, p_{i2}, \ldots, p_{i(K+1)})$ that are a function of $x_i$. The $K+1$ probabilities sum to one, so they can be parametrized by the vector $p_i = (p_{i1}, p_{i2}, \ldots, p_{iK})^T$.

Let $P$ be the length of $x_i$ and let $B$ be a $P \times K$ matrix of regression coefficients. Let $b_0$ be a vector of $K$ intercept values. The covariates are mapped to a vector of $K$ linear predictors, $\eta_i$, by the relationship $\eta_i = b_0 + B^T x_i$. Class probabilities are linked to the linear predictors by $\eta_i = g(p_i)$, where $g: \mathcal{S}^K \rightarrow \mathbb{R}^K$ is a multivariate invertible link function and $\mathcal{S}^K = \{p: p \in (0, 1)^K, \lVert p \rVert_1 <1 \}$. Furthermore, $g$ is a composite of two functions, $g_{EL}: (0, 1)^K \rightarrow \mathbb{R}^K$ and $g_{MO}: \mathcal{S}^K \rightarrow (0, 1)^K$. More specifically, ELMO class models have a link function of the form
\begin{equation*}
g(p) = (g_{EL} \circ g_{MO})(p)\ ,
\end{equation*}
where
\begin{equation*}
g_{MO}(p) = \delta = (\delta_1, \ldots, \delta_K)^T
\end{equation*}
and
\begin{equation*}
g_{EL}(\delta) = (g_{EL*}(\delta_1), \ldots, g_{EL*}(\delta_K))^T \ .
\end{equation*}

\subsection{Family function}\label{subsec:family}
The function $g_{MO}$ determines the family of multinomial-ordinal models, such as cumulative probability, stopping ratio, continuation ratio, or adjacent category. In order to belong to a multinomial-ordinal family, the function $g_{MO}$ must be invertible and have the following Monotonicity Property. This Monotonicity Property ensures that all parallel models in the ELMO class are in fact ordinal models (discussed in Section \ref{subsec:forms}). Examples of MO families are given in Table \ref{tab:mo}.

\noindent \textbf{Definition (Monotonicity Property) }
{\it For any $p \in \mathcal{S}^K$, define $\gamma_j(p)$ for $j = 1, \ldots, K$ as the sum of the first $j$ elements (i.e. cumulative probabilities). Define $\delta_i = (\delta_{i1}, \ldots, \delta_{iK})^T = g_{MO}(p_i)$ for $i \in \{1, 2\}$. All MO families have either Property 1 or Property 2 below.
\begin{enumerate}
	\item $\gamma_j(p_1) \le \gamma_j(p_2)$ for all $j$ if and only if $\delta_{1j} \le \delta_{2j}$ for all $j$.
	\item $\gamma_j(p_1) \le \gamma_j(p_2)$ for all $j$ if and only if $\delta_{1j} \ge \delta_{2j}$ for all $j$.
\end{enumerate}}



\begin{table}[H]
	\begin{center}
		\begin{tabular}{|c|c|}
			\hline
			MO Family & $\delta_j$ \\
			\hline
			Cumulative Probability (forward) & $\text{P}(Y \le j)$ \\
			Cumulative Probability (backward) & $\text{P}(Y \ge j+1)$ \\
			Stopping Ratio (forward) & $\text{P}(Y = j | Y \ge j)$ \\
			Stopping Ratio (backward) & $\text{P}(Y = j+1 | Y \le j+1)$ \\
			Continuation Ratio (forward) & $\text{P}(Y > j | Y \ge j)$ \\
			Continuation Ratio (backward) & $\text{P}(Y < j | Y \le j)$ \\
			Adjacent Category (forward) & $\text{P}(Y = j+1 | j \le Y \le j+1)$ \\
			Adjacent Category (backward) & $\text{P}(Y = j | j \le Y \le j+1)$ \\
			\hline
		\end{tabular}
	\end{center}
	\caption{Examples of multinomial-ordinal (MO) families. For each example, $Y$ is a categorical random variable with class probability vector $(p_1, p_2, \ldots, p_K) = g_{MO}^{-1}(\delta_1, \delta_2, \ldots, \delta_K)$.}
	\label{tab:mo}
\end{table}

Recall that $p_i = (p_{i1}, p_{i2}, \ldots, p_{iK})^T$, and let $r(p_i) = (p_{i(K+1)}, p_{iK}, \ldots, p_{i2})$ denote the class probabilities in reverse order, leaving out class 1 instead of class $K+1$. If $g_{MO}$ is a MO function with Property 1, then the $(g_{MO} \circ r)$ is a MO function with Property 2 and vice versa. We can refer to one as the ``forward'' family and the other as the ``backward'' family. Although the terms ``forward'' and ``backward'' are commonly used in the literature, they do not have a consistent interpretation. We follow the naming conventions used in \citet{Yee2010}. By these definitions, the forward cumulative probability and stopping ratio families have Property 1, and the forward continuation ratio and adjacent category families have Property 2.

In addition, if $g_{MO}$ is an MO function, then $g_{MO}^*(p) = 1 - g_{MO}(p)$ is also an MO function. For the cumulative probability and adjacent category families, this is simply a transformation between the forward and backward families. On the other hand, applying this transformation to the forward (backward) stopping ratio family yields the forward (backward) continuation ratio family.

\subsection{Elementwise link function}\label{subsec:element}
The elementwise link function $g_{EL*}: (0, 1) \rightarrow \mathbb{R}$ must be a monotone, invertible function. It can be almost any link function used for binary data regression, such as logit, probit, or complementary log-log. An important property that some elmentwise link functions satisfy is symmetry, that is
$$
g_{EL*}(\delta) = -g_{EL*}(1-\delta) \ .
$$
For example, logit and probit are symmetric, but complementary log-log is not. Under symmetry, the following model pairs are equivalent, with reversed signs on the coefficients: 1) cumulative probability forward and backward models; 2) adjacent category forward and backward models; 3)  forward stopping ratio and forward continuation ratio; and 4) backward stopping ratio and backward continuation ratio. However, if $g_{EL*}$ is not symmetric, such as the complementary log-log, then none of these equivalences hold.

\subsection{Parallel and nonparallel forms}\label{subsec:forms}
Each model in the ELMO class has a parallel form and a nonparallel form. The difference between them is that the parallel form restricts the columns of $B$ to be identical. This restriction is referred to as the parallelism assumption by \citet{Yee2010}. Let $b$ denote the common column vector and consider the distribution of $\bm{y}_i | \bm{x}_i=x_i$. The Monotonicity Property of $g_{MO}$, along with the monotonicity requirement on $g_{EL*}$, ensures that a change in $b^T x_i$ will shift all cumulative class probabilities in the same direction. This is the defining characteristic of an ordinal regression model.

In contrast, the nonparallel form places no restriction on $B$, and it does not force the cumulative class probabilities to ``shift together'' in any way. The nonparallel form is appropriate for unordered multinomial data, although it can also be used as a more flexible model for ordinal data.

A word of caution: the nonparallel cumulative probability model must a have linear predictor vector, $B^T x$, that is monotone to ensure that the cumulative probabilities are non-decreasing. For example, $B^T x$ must be monotone increasing for the forward model and monotone decreasing for the backward model. Thus, $B$ should be constrained such that $B^T x$ is monotone for any feasible $x$ in the population of interest. This constraint is difficult to implement in practice, especially because the range of feasible $x$ values may not be known. It is more practical to constrain $B^T x$ to be monotone for all $x$ in the training sample. However, this may lead to non-monotone probabilities for out-of-sample $x$, so it is important to be mindful of this. This is not a concern for the parallel cumulative probability model because the MLE (or penalized MLE) will always have monotone intercepts, and hence monotone linear predictors for all $x$. The other families in Table \ref{tab:mo} do not have any restriction on the parameter space.

\subsection{Semi-parallel form}\label{subsec:semi}
In most applications with ordinal response data, domain knowledge does not make it clear whether to use the parallel or nonparallel form. With enough observations, one could simply estimate the nonparallel model by maximum likelihood and obtain a good fit. After all, the nonparallel model includes the parallel model as a special case, and the parallel model will give inconsistent coefficient estimates if it is incorrect.

When the number of predictors is large relative to the number of observations, a regularization method such as lasso or elastic net is required. In this case, it is not possible to estimate each coefficient with a high degree of accuracy---a more realistic modeling goal is to build a model for out-of-sample prediction and determine the most important predictors. In this context, one might forgive some incorrectness of the parallel model if it is accurate enough to accomplish the modeling goals. Even if a nonparallel model were the true data generating mechanism, the regularized parallel model could still outperform the regularized nonparallel model for prediction.

The question becomes: how ``parallel'' does the data need to be to make the parallel model a better choice? If the response categories have a natural ordering, then it seems prudent to leverage this fact by using an ordinal regression model. However, the fact alone that the response is ordinal does not mean that a parallel regression model will be a good fit. Therefore, it also seems prudent to use a model that is sufficiently flexible to allow deviation from the strict parallelism assumption.

With this motivation, we propose a model that (1) is ordinal in nature and (2) allows deviation from the parallelism assumption. We call this the \emph{semi-parallel} model. Recall that the nonparallel model specifies $\eta_i = b_0 + B^T x_i$, where $b_0$ is a vector of $K$ intercepts and $B$ is an unrestricted $P \times K$ matrix of coefficients. The parallel model restricts the columns of $B$ to be identical and can be parametrized as $\eta_i = b_0 + (b^T x_i) \cdot \mathbbm{1}$. The semi-parallel model specifies $\eta_i = b_0 + B^T x_i +( b^T x_i) \cdot \mathbbm{1}$. It is the nonparallel model but overparametrized to include both the parallel and nonparallel coefficients. With the elastic net penalty, the penalized likelihood has a unique solution in most cases (see Appendix \ref{app:uniqueness} for details). We use the term \emph{semi-parallel} because for some covariates, the penalized semi-parallel model fit might only contain the parallel coefficient, with the nonparallel coefficients all set to zero. For other covariates, the fit might contain both parallel and nonparallel coefficients.

\subsection{Elastic net penalty}\label{subsec:ENP}
This section discusses the elastic net penalty for ELMO models. There are many useful resources on  regularization,  penalized regression, and variable selection, which provide further details in various settings \citep{Hastie2009, bickel2006regularization, hesterberg2008least, Friedman2010, schifano2010majorization, vidaurre2013survey}.

If the sample size is large enough, it may be possible to accurately estimate a regression model by maximum likelihood. However, in many applications the sample size is not large enough to obtain reliable or even unique estimates. In situations like this, it may be advantageous to optimize a penalized version of the log likelihood function. One such penalty is the elastic net \citep{Zou2005}, which is a generalization of both the lasso \citep{Tibshirani1996} and ridge regression penalties.

Lasso and ridge regression are techniques that minimize a penalized likelihood objective function, defined as the negative log-likelihood plus a penalty term that is a function of the coefficient vector. For lasso, the penalty is proportional to the $L_1$ norm of the coefficient vector, and for ridge regression it is proportional to the squared $L_2$ norm. Both of these penalties result in a coefficient estimate that is closer to zero than the maximum likelihood estimator, i.e. the estimate is ``shrunk'' toward zero. This biases the estimates toward zero, but the trade-off is a reduction in variance which often reduces the overall mean squared error. The lasso also has the property that some of the coefficient estimates are shrunk to zero exactly. This provides a natural way to perform variable selection because only the the predictors most associated with the response variable will have nonzero coefficients.

The elastic net penalty is a weighted average between the lasso and ridge regression penalties, and it shares the lasso property of shrinking some coefficients to zero exactly. The weighting parameter, typically denoted by $\alpha$, must be selected or tuned on the data set. The degree of penalization is controlled by another tuning parameter, typically denoted by $\lambda$. Typical practice is to fit the penalized model for a sequence of $\lambda$ values and use a tuning procedure to select the best value \citep{hesterberg2008least, Hastie2009, arlot2010survey, sun2013consistent}. One tuning procedure is to select the model with best fit as determined by $C_p$, AIC, BIC, or another fitness measure. An alternative approach is using cross-validation to select the value that gives the best out-of-sample prediction. We use cross-validation.

Let $b_j$ be the $j^\text{th}$ element of $b$ and $B_{jk}$ be the element in the $j^\text{th}$ row and $k^\text{th}$ column of $B$. Let $N_* = \sum_{i=1}^N n_i$ be the total number of multinomial trials in the data set. Let $\ell(\cdot)$ denote the log-likelihood function for each ELMO model form. Below, we write the elastic net objective function for each model form.

\subsubsection*{Parallel model}
The objective function is
\begin{equation*}
\mathcal{M}(b_0, b; \alpha, \lambda) = -\frac{1}{N_*} \ell(b_0, b) + \lambda \sum\limits_{j=1}^P \left( \alpha |b_j| + \tfrac{1}{2} (1-\alpha) b_j^2 \right) \ .
\end{equation*}

\subsubsection*{Nonparallel model}
The objective function is
\begin{equation*}
\mathcal{M}(b_0, B; \alpha, \lambda) = -\frac{1}{N_*} \ell(b_0, B) + \lambda \sum\limits_{j=1}^P \sum\limits_{k=1}^K \left( \alpha |B_{jk}| + \tfrac{1}{2} (1-\alpha) B_{jk}^2 \right) \ .
\end{equation*}

\subsubsection*{Semi-parallel model}
The objective function is
\begin{equation*}
\begin{split}
&\mathcal{M}(b_0, b, B; \alpha, \lambda, \rho) = -\frac{1}{N_*} \ell(b_0, b, B) + \\
& \quad + \lambda \left( \rho \sum\limits_{j=1}^P \left( \alpha |b_j| + \tfrac{1}{2} (1-\alpha) b_j^2 \right) + \sum\limits_{j=1}^P \sum\limits_{k=1}^K \left( \alpha |B_{jk}| + \tfrac{1}{2} (1-\alpha) B_{jk}^2 \right) \right) \ .
\end{split}
\end{equation*}

Here, $\lambda \ge 0$ and $\alpha \in [0, 1]$ are the tuning parameters previously described. Also, $\rho \ge 0$ is a tuning parameter that determines the degree to which the parallel terms are penalized.  Fixing $\rho$ at a very large value will set all parallel coefficients to zero, which is equivalent to the nonparallel model. Fixing $\lambda$ at a very large value and choosing $\rho$ such that $\lambda \rho = \lambda^*$ is equivalent to the parallel model with regularization parameter $\lambda^*$. Hence, the semi-parallel model includes both the parallel and nonparallel models as special cases. Fixing $\rho = 0$ will leave the parallel coefficients unpenalized, so the fit will shrink from the maximum likelihood nonparallel model fit toward the maximum likelihood parallel model fit as $\lambda$ increases from zero.

We follow the common convention of scaling the negative log-likelihood by the number of observations \citep{hesterberg2008least}. This way, when fitting a model to a sample from a given population, a given $\lambda$ value will have roughly the same degree of penalization regardless of the sample size. This is convenient when tuning $\lambda$ by cross validation because the tuning data set may have a different sample size than the training data set. We define the sample size as $N_*$ rather than $N$ so the model fit is invariant to whether multinomial trials are grouped into a single observation or split into multiple observations with $n_i = 1$ and identical $x$.

%
%
%
%

\subsection{Inverse link function Jacobian}\label{subsec:ILFJ}
The Jacobian of the inverse link function is required for the coordinate descent algorithm. This computation can be compartmentalized for link functions in the ELMO class because of their composite form. Define $h$, $h_{EL}$, $h_{EL*}$, and $h_{MO}$ to be the inverses of $g$, $g_{EL}$, $g_{EL*}$, and $g_{MO}$, respectively. The inverse link function can be written as
\begin{equation*}
h(\eta) = h_{MO}(\delta) = h_{MO}(h_{EL}(\eta)) \ ,
\end{equation*}
where $h_{EL}(\eta) = \left( h_{EL*}(\eta_1),\ h_{EL*}(\eta_2),\ \ldots,\ h_{EL*}(\eta_K) \right)$.

The Jacobian of the inverse link can be written as
\begin{equation*}
D h(\eta) = D h_{EL}(\eta) \ D h_{MO}(p) \ ,
\end{equation*}
where $D h_{EL}(\eta) = \text{diag} \left\{ h_{EL*}'(\eta_1),\ h_{EL*}'(\eta_2), \ldots,\ h_{EL*}'(\eta_K) \right\}.$

The inverse and its derivative are well-known for common elementwise link functions, so we do not discuss these any further (see, e.g., the \texttt{make.link} function in the R package \textbf{stats}). To calculate $h(\eta)$, it only remains to calculate $h_{MO}(\delta)$ and $D h_{MO}(\delta)$. These calculations are presented for specific MO families in Appendix \ref{app:mo}.


\section{Coordinate descent optimization algorithm}\label{sec:cd}
We propose optimizing ELMO class models with the elastic net penalty using a coordinate descent algorithm. Our algorithm mirrors that of \citet{Friedman2007, Friedman2010} for generalized linear models. The algorithm is iterative and has an outer and inner loop. The outer loop constructs a quadratic approximation to the log-likelihood---the same quadratic approximation used for iteratively reweighted least squares (IRLS). This approximation is the second order Taylor expansion at the current coefficient estimates. In the spirit of Fisher scoring, the Hessian matrix is replaced by its expectation, the negative Fisher information matrix. This approximation is used as a replacement for the true likelihood in the elastic net objective function, resulting in an expression that can be optimized by coordinate descent. The inner loop cycles through the coefficient estimates, updating each one with the value that marginally optimizes the approximate objective function.

\citet{Wilhelm1998} demonstrated the use of IRLS to obtain the maximum likelihood estimates for a broad class of multinomial regression models. This class includes ELMO models but is even more general. This algorithm can easily be applied to any multinomial regression model that links a vector of $K$ probabilities to a vector of $K$ linear combinations of covariates. The link function $g$ does not need to be a composite function or have the Monotonicity Property of the ELMO class. It simply needs to have an inverse $h$. To apply the coordinate descent algorithm to another model, all that is required is to derive the Jacobian of $h$. Although ELMO is a rich class of models, there are multinomial regression models outside this class (e.g. multinomial logistic regression).  The coordinate descent algorithm is very general. One could use the basic ideas for fitting an elastic net penalized multinomial regression model that is outside the ELMO class.

The rest of this section is organized as follows. In order to formulate the IRLS quadratic approximation, it is more convenient to parametrize ELMO models with a single coefficient vector instead of $b_0$, $b$, and $B$. Section \ref{subsec:elmo-single} discusses this alternative parameterization.   Section \ref{sec:enet-optim} discusses the elastic net penalty under the alternative parametrization. In Section \ref{sec:scoreinfo} we derive the general form of the score and information matrix for multinomial regression models. Section \ref{sec:cd-outer} discusses the outer loop of the optimization procedure, which updates the quadratic approximation to the log-likelihood. Section \ref{sec:cd-inner} discusses the coordinate descent inner loop. Section \ref{subsec:numerical} discusses computational efficiency and numerical stability for the coordinate descent updates.

Sections \ref{subsec:lambda}, \ref{subsec:start}, and  \ref{subsec:stop} discuss different aspects of the algorithm specifications. Specifically, Section  \ref{subsec:lambda} presents a method for choosing a sequence of regularization parameter values for the solution path. Section  \ref{subsec:start} presents a method for choosing starting coefficient values for the optimization algorithm. And Section  \ref{subsec:stop} suggests a stopping rule for terminating the algorithm. Section \ref{subsec:summary} summarizes the  algorithm in outline form. Finally, Section \ref{subsec:cp-constraint} discusses specific optimization issues that can arise with the cumulative probability model family.

\subsection{ELMO parametrization with a single coefficient vector}\label{subsec:elmo-single}
Until now, we have written ELMO coefficients in a compact form using an intercept vector $b_0$, a coefficient vector $b$, and a coefficient matrix $B$. For coordinate descent, it is more convenient to write the model with a single coefficient vector $\beta$. To do this, we need to introduce a covariate matrix $X_i$, which is a function of $x_i$. The vector of $K$ linear combinations, $\eta_i$, can then be written as $\eta_i = X_i^T \beta$ for any of the three ELMO model forms.

Let $B_{j \cdot}$ denote the transpose of the $j^\text{th}$ row of $B$. The form of $X_i$ and $\beta$ is given below for the parallel, nonparallel, and semi-parallel models.

\subsubsection*{Parallel model}
\begin{equation*}
X_i = \left(\begin{array}{c|c}
	I_{K \times K} & \begin{array}{ccc}
		x_i^T \\
		\vdots \\
		x_i^T
	\end{array}
\end{array} \right)_{K \times (P+K)},
\hspace{2em}
\beta = \begin{pmatrix}
	b_0 \\
	b
\end{pmatrix}_{(P+K) \times 1}.
\end{equation*}

\subsubsection*{Nonparallel model}
\begin{equation*}
X_i = \left(\begin{array}{c|c}
	I_{K \times K} & \begin{array}{cccc}
		x_i^T & 0 & \cdots & 0 \\
		0 & x_i^T & \cdots & 0 \\
		\vdots & \vdots & \ddots & \vdots \\
		0 & 0 & \cdots & x_i^T
	\end{array}
\end{array} \right)_{K \times (PK+K)},
\hspace{2em}
\beta = \begin{pmatrix}
	b_0 \\
	B_{1 \cdot} \\
	B_{2 \cdot} \\
	\vdots \\
	B_{K \cdot}
\end{pmatrix}_{(PK+K) \times 1}.
\end{equation*}

\subsubsection*{Semi-parallel model}
\begin{equation*}
X_i = \left(\begin{array}{c|c}
	I_{K \times K} & \begin{array}{ccccc}
		x_i^T & x_i^T & 0 & \cdots & 0  \\
		x_i^T & 0 & x_i^T & \cdots & 0 \\
		\vdots &  \vdots & \vdots & \ddots & \vdots \\
		x_i^T & 0 & 0 & \cdots & x_i^T
	\end{array}
\end{array} \right)_{K \times (P(K+1)+K)},
\hspace{2em}
\beta = \begin{pmatrix}
	b_0 \\
	b \\
	B_{1 \cdot} \\
	B_{2 \cdot} \\
	\vdots \\
	B_{K \cdot}
\end{pmatrix}_{(P[K+1]+K) \times 1}.
\end{equation*}

\subsection{Elastic net penalty}\label{sec:enet-optim}

Suppose $\beta$ is length $Q$ and let $\beta_j$ denote the $j^\text{th}$ element. We write the elastic net objective function as
\begin{equation*}
\mathcal{M}(\beta; \alpha, \lambda, c_1, \ldots, c_Q) = -\frac{1}{N_*} \ell(\beta) + \lambda \sum\limits_{j=1}^{Q} c_j \left( \alpha |\beta_j| + \tfrac{1}{2} (1-\alpha) \beta_j^2 \right) \ ,
\end{equation*}
where $\lambda > 0$ and $0 \le \alpha \le 1$. For ELMO models, $c_j = 0$ for all intercept terms. For the parallel and nonparallel models, $c_j = 1$ for all non-intercept coefficients. For the semi-parallel model, $c_j = \rho$ for the parallel non-intercepts ($b$), and $c_j = 1$ for the nonparallel non-intercepts ($B$). These are not firm rules regarding $c_j$, as there may be situations where one wishes to modify the $c_j$ to accommodate more elaborate penalization schemes. For example, one might wish leave to some covariates unpenalized or to penalize them with varying degrees. The only requirement is that the $c_j$ be nonnegative.

Typically, each covariate is standardized to have its sample standard deviation equal to one so that the scale of a covariate does not affect the degree to which its coefficient is penalized \citep{hesterberg2008least}. However, this is not a requirement.

\subsection{Score and information matrix}\label{sec:scoreinfo}

In this section, we derive the general form of the score and information matrix for multinomial regression models. The log-likelihood of an observation with probability vector $p_i$ can be written as
\begin{equation*}
	L_i(p_i) = \sum\limits_{j=1}^K y_{ij} \log(p_{ij}) + y_{i(K+1)}\log\left(1-\sum\limits_{j=1}^K p_{ij}\right) \ .
\end{equation*}
Note that we drop the multinomial term $\log \binom{n_i}{y{i1}, y_{i2}, \ldots, y_{i(K+1)}}$ because it does not depend on the unknown coefficients, and hence does not affect the model fit. The log-likelihood as a function of $\beta$ is
\begin{equation*}
	\ell_i(\beta) = L_i(h(X_i^T \beta)) \ .
\end{equation*}

The score function can be obtained by a chain rule decomposition:
\begin{equation*}
	U_i(\beta) = X_i^T D h(\eta_i)^T \nabla L_i(p_i) = X_i^T v_i \ ,
\end{equation*}
\begin{equation*}
	\begin{split}
		\text{where } & D h(\eta_i) = \left( \frac{\partial h_1}{\partial \eta_i}, \ldots, \frac{\partial h_K}{\partial \eta_i} \right) \ , \\
		& \nabla L_i(p_i) = \left( \frac{y_{i1}}{p_{i1}}, \ldots, \frac{y_{iK}}{p_{iK}} \right)^T - \left( \frac{y_{i(K+1)}}{p_{i(K+1)}} \right) \cdot \mathbbm{1} \text{ , and} \\
	    & v_i = D h(\eta_i)^T \nabla L_i(p_i) \ .
\end{split}
\end{equation*}

The Fisher information matrix is given by
\begin{equation*}
	\mathcal{I}_i(\beta) = E_{\beta} \{ U_i(\beta) U_i(\beta)^T \} = X_i^T W_i X_i \ ,
\end{equation*}
where $W_i = D h(\eta_i)^T \Sigma_i^{-1} D h(\eta_i)$ and
\begin{equation*}
	\begin{split}
		\Sigma_i^{-1}
		&= E_{\beta} \left\{ \nabla L_i(p_i) \nabla L_i(p_i)^T \right\} \\
		&= \left[ E_{\beta} \left\{ \left( \frac{y_{im}}{p_{im}} - \frac{y_{i(K+1)}}{p_{i(K+1)}} \right)
		\left( \frac{y_{in}}{p_{in}} - \frac{y_{i(K+1)}}{p_{i(K+1)}} \right) \right\} \right]_{mn} \\
		&= \left[ n_i \left( \frac{I(m=n)}{p_{im}} + \frac{1}{p_{i(K+1)}} \right) \right]_{mn} \\
		&= n_i \left( \left[ \text{diag}(p_i) \right]^{-1} + \frac{1}{p_{i(K+1)}} \cdot \mathbbm{1} \mathbbm{1}^T \right)\ .
	\end{split}
\end{equation*}

Because the $y_i$ are independent, the full data log-likelihood, score, and information are defined as

\begin{equation*}
	\begin{split}
		& \ell(\beta) = \sum_{i=1}^N \ell_i(\beta) \ , \\
		& U(\beta) = \sum_{i=1}^N U_i(\beta) = X^T v \text{ , and} \\
		& \mathcal{I}(\beta) = \sum_{i=1}^N \mathcal{I}_i(\beta) = X^T W X \ ,
	\end{split}
\end{equation*}

\begin{equation*}
\text{where }
X = \begin{pmatrix}
	X_1 \\
	\vdots \\
	X_N
\end{pmatrix} \ , \ \
v = \begin{pmatrix}
	v_1 \\
	\vdots \\
	v_N
\end{pmatrix} \ , \ \
\text{and \ }
W =  \begin{pmatrix}
	W_1 & 0 & \cdots & 0 \\
	0 & W_2 & \cdots & 0 \\
	\vdots & \vdots & \ddots & \vdots \\
	0 & 0 & \cdots & W_N
\end{pmatrix} \ .
\end{equation*}

\subsection{Optimization outer loop (quadratic approximation)}\label{sec:cd-outer}
The optimization outer loop updates the quadratic approximation to the log-likelihood using a Taylor expansion around the current coefficient estimates. Let $\hat{\beta}^{(r)}$ denote the coefficient estimates after the $r^{\text{th}}$ outer loop iteration, and let the $(r)$ superscript denote terms that depend on $\hat{\beta}^{(r)}$. Of course, starting values are required for the first iteration, and this topic will be discussed later.

We define the quadratic approximation of $\ell(\beta)$ as
\begin{equation}
\ell^{(r)}(\beta) = -\tfrac{1}{2} (z^{(r)} - X \beta)^T\ W^{(r)}\ (z^{(r)} - X \beta) \ ,
\end{equation}
where $z^{(r)} = X \hat{\beta}^{(r)} + \{W^{(r)}\}^{-1} v^{(r)}$. This is the second order Taylor series expansion of $\ell$ at $\hat{\beta}^{(r)}$, up to an additive constant that does not depend on $\beta$. Also, the Hessian is replaced by its expectation, $-\mathcal{I}(\hat{\beta}^{(r)})$. The derivation is provided in Appendix \ref{sec:quadapp}. The inner loop computes $\hat{\beta}^{(r+1)}$ by optimizing the \textit{penalized} quadratic approximation.

\subsection{Optimization inner loop (coordinate descent)}\label{sec:cd-inner}
For unpenalized maximum likelihood estimation, the quadratic approximation is maximized by the usual weighted least squares solution $\hat{\beta}^{(r+1)} = (X^T W^{(r)} X)^{-1} X^T W^{(r)} z^{(r)}$. With the elastic net penalty, we can still follow the IRLS procedure, but the optimization step no longer has a closed form solution. This is because partial derivatives of the elastic net penalty do not exist at zero for any of the penalized coefficients. The optimization step can instead be done with a coordinate descent procedure. This involves cycling through the coefficient estimates, updating each one with the value that marginally optimizes the approximate objective function. The cycle is iterated until convergence.

Let $\mathcal{M}^{(r)}(\beta)$ denote the elastic net objective function with $\ell^{(r)}(\beta)$ in place of $\ell(\beta)$. Let $\hat{\beta}_j^{(r, s)}$ denote the estimate of $\beta_j$ at the $s^\text{th}$ inner loop iteration of the $r^\text{th}$ outer loop iteration. Let	$\mathcal{M}_{j}^{(r, s)}(t) = \mathcal{M}^{(r)}(\hat{\beta}_1^{(r, s+1)}, \ldots, \hat{\beta}_{j-1}^{(r, s+1)}, t, \hat{\beta}_{j+1}^{(r, s)}, \ldots, \hat{\beta}_{Q}^{(r, s)})$ denote $\mathcal{M}^{(r)}$ as a marginal function of the $j^\text{th}$ regression coefficient only, with all other coefficients fixed at their current estimates. The $s^\text{th}$ iteration coordinate descent update of the $j^\text{th}$ coefficient is $\arg\min \mathcal{M}_j^{(r, s)}(t)$. If $c_j=0$ (i.e. $\beta_j$ is unpenalized), then this can be solved straightforwardly by setting $\tfrac{d}{dt} \mathcal{M}_j^{(r, s)}(t) = 0$. In general, for $t \ne 0$,
\begin{equation*}
\frac{d}{dt} \mathcal{M}_j^{(r, s)}(t) = -\frac{1}{N_*} X_{\cdot j}^{T} W^{(r)} \left( z^{(r)} - X_{\cdot -j} \hat{\beta}_{-j}^{(r, s)} - t X_{\cdot j} \right) + \lambda \left( \alpha \cdot \text{sign}(t) + (1-\alpha) \cdot t \right) \ ,
\end{equation*}
where $X_{\cdot j}$ denotes the $j^\text{th}$ column of $X$, $X_{\cdot -j}$ denotes $X$ with the $j^\text{th}$ column deleted, and $\hat{\beta}_{-j}^{(r, s)} = (\hat{\beta}_1^{(r, s+1)}, \ldots, \hat{\beta}_{j-1}^{(r, s+1)}, \hat{\beta}_{j+1}^{(r, s)}, \ldots, \hat{\beta}_{Q}^{(r, s)})$. $\mathcal{M}_j^{(r, s)}(t)$ is convex because $\tfrac{d}{dt} \mathcal{M}_j^{(r, s)}(t)$ runs from $-\infty$ to $+\infty$ and is monotonically increasing. The only point where the derivative does not exist is at $t = 0$, where it jumps up by $2 \lambda \alpha c_j$. If $\left| \tfrac{1}{N_*} X_{\cdot j}^{T} W^{(r)} \left(z^{(r)} - X_{\cdot -j} \hat{\beta}_{-j} \right) \right| > \lambda \alpha c_j$, then the derivative attains zero, and the value at which this occurs is $\arg \min \mathcal{M}_j^{(r, s)}(t)$. Otherwise the derivative changes sign at $t = 0$, and $\arg \min \mathcal{M}_j^{(r, s)}(t) = 0$. Hence, the coordinate descent update can be written as
\begin{equation}
\label{eqn:cdupdate}
\hat{\beta}_j^{(r, s+1)} = \frac{S\left(\tfrac{1}{N_*} X_{\cdot j}^T W^{(r)} \left(z^{(r)} - X_{\cdot -j} \hat{\beta}_{-j}^{(r, s)} \right),\ \lambda \alpha c_j \right)}{\tfrac{1}{N_*} X_{\cdot j}^T W^{(r)} X_{\cdot j} + \lambda(1-\alpha) c_j} \ ,
\end{equation}
where $S(x, y) = \text{sign}(x)(|x| - y)_+$ is the soft-thresholding operator, as defined in \citet{Friedman2010}.

On a side note, in some situations one may wish to include a model constraint that forces some or all of the coefficients to be nonnegative (e.g. to implement the nonnegative garrote penalty discussed in \citet{Breiman1995}). With this constraint, the coordinate descent update becomes $\hat{\beta}_j^{(r, s+1)}  = \underset{t \ge 0}{\arg \min} \mathcal{M}_j^{(r, s)}(t)$, where the only difference is the restriction that $t \ge 0$. When $\arg \min \mathcal{M}_j^{(r, s)}(t) < 0$, then $\underset{t \ge 0}{\arg \min} \mathcal{M}_j^{(r, s)}(t) = 0$ because $\mathcal{M}_j^{(r, s)}(t)$ is convex. The coordinate descent update can be written as
\begin{equation*}
\hat{\beta}_j^{(r, s+1)} = \frac{\left(\tfrac{1}{N_*} X_{\cdot j}^T W^{(r)} \left(z^{(r)} - X_{\cdot -j} \hat{\beta}_{-j}^{(r, s)} \right) - \lambda \alpha c_j \right)_+}{\tfrac{1}{N_*} X_{\cdot j}^T W^{(r)} X_{\cdot j} + \lambda(1-\alpha) c_j} \ .
\end{equation*}

\subsection{Computational efficiency and numerical stability}\label{subsec:numerical}
It would be computationally inefficient to compute the coordinate descent coefficient updates in the form presented in Section \ref{sec:cd-inner}. First of all, $W^{(r)}$ is a sparse block diagonal matrix and should not be multiplied without taking this into consideration. Secondly, it is not necessary to explicitly compute $z^{(r)}$, which depends on $W_1^{-1}$, $W_2^{-1}$, \ldots, $W_N^{-1}$. It is not necessarily computationally expensive to compute these $K \times K$ matrix inverses when $K$ is small, but it is still better to avoid doing so. For computational efficiency, it is better to write the coordinate descent updates in terms of the score and information matrix. This is done by making the following two substitutions in \eqref{eqn:cdupdate}
\begin{eqnarray*}
X_{\cdot j}^T W^{(r)} \left(z^{(r)} - X_{\cdot -j} \hat{\beta}_{-j}^{(r, s)} \right) &=& \left[ U(\hat{\beta}^{(r)}) \right]_j + \left[\mathcal{I}(\hat{\beta}^{(r)}) \hat{\beta}^{(r)} \right]_j - \left[ \mathcal{I}(\hat{\beta}^{(r)}) \hat{\beta}^{(r, s)} \right]_j, \\
X_{\cdot j}^T W^{(r)} X_{\cdot j} &=& \left[ \mathcal{I}(\hat{\beta}^{(r)}) \right]_{jj}.
\end{eqnarray*}

The subscripts on terms with square brackets indicate the $j^\text{th}$ vector element or matrix diagonal. Only the third term on the right hand side of the first substitution needs to be updated with each inner loop iteration because it depends on $\hat{\beta}^{(r, s)}$. The other terms are a function of $\hat{\beta}^{(r)}$, so they only need to be updated during the outer loop.

Furthermore, the $X_i$ matrices are sparse for ELMO models. In some programming languages, it may be advantageous to use this block structure to compute $X_i \hat{\beta}$ and $X_i^T W_i X_i$. This has the additional advantage that it is not necessary to store $X_1$, $X_2$, \ldots, $X_N$ in memory, but rather just $x_1$, $x_2$, \ldots, $x_N$, which are smaller.

Numerical instability of the information matrix can arise when the fitted class probabilities approach zero for any observation. This is problematic even if the near-zero probability occurs for an observation $i$ and class $j$ such that $y_{ij} = 0$. A way to prevent numerical instability is to cap the fitted probabilities at some minimum value just for the information matrix calculation. (The \texttt{pMin} argument sets this threshold for the optimization function \texttt{ordinalNet} in the \textbf{ordinalNet} R package.) The score and likelihood are computed with uncapped fitted probabilities.

\subsection{Regularization parameter sequence}\label{subsec:lambda}
Often we are interested in computing solutions for a sequence of $\lambda$ values, rather than a single value. For $\alpha > 0$, there always exists a threshold value $\lambda_{\max}$ where the first coefficient enters the solution path. All penalized coefficients are set to zero for any $\lambda > \lambda_{\max}$. An off-the-shelf method to generate a reasonable sequence of $\lambda$ values is to let $\lambda_{\min} = 0.01 \times \lambda_{\max}$ and consider a sequence of $\lambda$ values that is uniform between $\lambda_{\max}$ and $\lambda_{\min}$ on the log scale \citep{Friedman2010}.

To calculate $\lambda_{\max}$, we first fit the intercept-only model by unpenalized maximum likelihood. Also include any unpenalized non-intercept coefficients if there are any. We then calculate the quadratic approximation at this solution. Each penalized coefficient has a threshold value of $\lambda$ where its coordinate descent update becomes nonzero. The minimum threshold value among all coefficients is $\lambda_{\max}$, as this is the value where the first coefficient enters the solution path. Specifically,
\begin{equation*}
	\lambda_{\max} = \underset{j}{\min} \frac{1}{N_* \alpha c_j} \left| X_{\cdot j}^T W \left(z - X_{\cdot -j} \hat{\beta}_{-j} \right) \right| \ ,
\end{equation*}
where $\hat{\beta}$ is the intercept-only maximum likelihood estimate, and $W$ and $z$ are calculated at $\hat{\beta}$.

\subsection{Starting values}\label{subsec:start}

\citet{Park2007} proposed an efficient estimation procedure for a decreasing sequence of $\lambda$ values. The $\hat{\beta}$ solution for each $\lambda$ value is used as the starting value for the next $\lambda$ in the sequence. This technique is known as ``warm starts.'' Furthermore, it is not necessary to update all coefficient estimates during the coordinate descent inner loop. Many coefficient estimates will begin and remain at zero throughout the inner loop, especially for larger $\lambda$ values. It is more efficient to cycle through only the coefficients that had nonzero estimates at the previous step. The set of nonzero coefficients is known as the ``active set.'' After the coordinate descent inner loop converges, one pass can be made over each coefficient outside the active set. If the coordinate descent update is zero for all of them, then the optimal solution has been reached. If the final pass changes any coefficient estimate from zero to a nonzero value, then the coordinate descent loop should be continued including these new nonzero coefficients in the active set.

A reasonable set of starting values can usually be obtained by passing the observed response category frequencies into the link function---this provides intercept starting values, and all other coefficients can start at zero. This is also the solution corresponding to $\lambda_{\max}$ if there are no unpenalized non-intercept coefficients. If the first $\lambda$ value is not $\lambda_{\max}$ or some of the non-intercepts are unpenalized, then this is still usually a reasonable set of starting values for the first $\lambda$ value.

\subsection{Stopping rule}\label{subsec:stop}

The coordinate descent procedure has an inner and outer loop, both of which require convergence definitions and thresholds. A suggestion is to define convergence using the relative change in the elastic net objective. For the outer loop, the definition is $\left| \frac{\mathcal{M}(\hat{\beta}^{(r)}) - \mathcal{M}(\hat{\beta}^{(r-1)})}{\mathcal{M}(\hat{\beta}^{(r-1)})} \right| < \epsilon_\text{out}$. For the inner loop, the quadratic approximation to the log-likelihood should be used instead of the true log-likelihood, so the definition is $\left| \frac{\mathcal{M}^{(r)}(\hat{\beta}^{(r, s)}) - \mathcal{M}^{(r)}(\hat{\beta}^{(r, s-1)})}{\mathcal{M}^{(r)}(\hat{\beta}^{(r, s-1)})} \right| < \epsilon_\text{in}$. A small constant can also be added to the denominator to allow convergence when the log-likelihood is near zero. Based on some trial and error, it seems efficient to set the outer and inner convergence thresholds, $\epsilon_\text{out}$ and $\epsilon_{\text{in}}$, to the same value.

\subsection{Algorithm summary}\label{subsec:summary}

\begin{enumerate}
	\item Fit the intercept-only model by maximum likelihood. (Also include any unpenalized non-intercept coefficients if there are any.)
	\item Calculate $\lambda_{\max}$ and choose a decreasing sequence $\lambda_{\max} = \lambda_1, \lambda_2, \ldots, \lambda_M = \lambda_{\min}$.
	\item Set $\hat{\beta}[\lambda_1]$ equal the solution of the intercept-only model.
	\item For $m=2$ to $M$:
	\begin{enumerate}
		\item Set $r \leftarrow 0$ and $\hat{\beta}^{(0)} \leftarrow \hat{\beta}[\lambda_{m-1}]$.
		\item While $\left| \frac{\mathcal{M}(\hat{\beta}^{(r)}) - \mathcal{M}(\hat{\beta}^{(r-1)})}{\mathcal{M}(\hat{\beta}^{(r-1)})} \right| > \epsilon_\text{out}$:
		\begin{enumerate}
			\item Calculate $U(\hat{\beta}^{(r)})$ and $\mathcal{I}(\hat{\beta}^{(r)})$.
			\item Set $s \leftarrow 0$ and $\hat{\beta}^{(r,0)} \leftarrow \hat{\beta}^{(r)}$.
			\item While $\left| \frac{\mathcal{M}^{(r)}(\hat{\beta}^{(r, s)}) - \mathcal{M}^{(r)}(\hat{\beta}^{(r, s-1)})}{\mathcal{M}^{(r)}(\hat{\beta}^{(r, s-1)})} \right| > \epsilon_\text{in}$:
			\begin{enumerate}
				\item Calculate $\hat{\beta}^{(r,s+1)}$ with a single cycle of coordinate descent updates over the coefficient active set.
				\item Set $s \leftarrow s+1$.
			\end{enumerate}
			\item Do one loop of coordinate descent updates over coefficients outside the active set. If any coefficient estimate changes to a nonzero value, then repeat the previous loop with the new nonzero coefficients in the active set.
			\item Set $\hat{\beta}^{(r+1)} \leftarrow \hat{\beta}^{(r,s)}$.
			\item Set $r \leftarrow r+1$.
		\end{enumerate}
	\item Set $\hat{\beta}[\lambda_m] \leftarrow \hat{\beta}^{(r)}$.
	\end{enumerate}
\end{enumerate}

\subsection{Issues with the cumulative probability family}\label{subsec:cp-constraint}

The cumulative probability family has a constrained parameter space because the cumulative probabilities must be monotone for every $x$ in the population. It was discussed in Section \ref{subsec:forms} that if the constraint is only enforced for $x$ in the training sample, then difficulties may arise because an out-of-sample $x$ may have fitted probabilities that are not monotone.

The parameter space constraint can also create difficulties during optimization. Although the likelihood is undefined outside the constrained parameter space, we could define an improper likelihood on the unrestricted parameter space. This improper likelihood would allow observations to have fitted probabilities greater than zero or less than one, and it would be defined as zero anywhere that an observation has negative probability in a class that was observed. This is essentially the likelihood that coordinate descent algorithm is designed to optimize. As a result, the algorithm may seek a solution that lies outside the constrained parameter space.

When the solution path leaves the constrained parameter space, we simply terminate the optimization algorithm at the $\lambda$ value where this occurred. Further work is required to devise a constrained optimization procedure for the cumulative probability family.


\section{Simulation}\label{sec:sim}
We have discussed three penalized ELMO model forms for ordinal data: the parallel model, the nonparallel model, and the semi-parallel model. The purpose of the following simulation experiments is to show scenarios where each of these three model types yields better out-of-sample prediction accuracy than the others.

We conducted three simulation experiments, each based on 100 replicates, i.e. simulated datasets.  For each replicate, the data were simulated from a forward stopping ratio model with the parameters shown in Table \ref{tab:sim-settings}. In all of the experiments, the covariates were simulated as independent, standard normal random variables.

Now consider the estimation procedures. For each of the three models, parallel, nonparallel, and semi-parallel, the elastic net tuning parameter was set to $\alpha = 1$ (i.e. lasso penalty). For the semi-parallel model, the tuning parameter $\rho$ was set to one. A $\lambda$ sequence of twenty values was generated uniformly on the log scale between $\lambda_{\max}$ of the full training data and $0.01 \times \lambda_{\max}$. For each simulated dataset, five-fold cross validation was used to select the optimal tuning parameter.

\begin{table}[h]
	\begin{center}
	\begin{tabular}{cccc}
		\hline
		& $N$ & $b_0$ & $B$ \\
		\hline
		Sim 1 & 500 & $\begin{pmatrix} -0.5 \\ 0 \end{pmatrix}$ &
		$\begin{pmatrix}
			0 & 2
		\end{pmatrix}$ \\
		\hline
		Sim 2 & 50 & $\begin{pmatrix} -0.5 \\ 0 \end{pmatrix}$ &
		$\begin{pmatrix}
			2 & 2 & \\
			\vdots & \vdots & \Big\} \times 5 \\
			2 & 2 & \\
			0 & 0 & \\
			\vdots & \vdots & \Big\} \times 10 \\
			0 & 0 &
		\end{pmatrix}$ \\
		\hline
		Sim 3 & 50 & $\begin{pmatrix} -0.5 \\ 0 \end{pmatrix}$ &
		$\begin{pmatrix}
		-2 & 2 & \\
		2 & 2 & \\
		\vdots & \vdots & \Big\} \times 4 \\
		2 & 2 & \\
		0 & 0 & \\
		\vdots & \vdots & \Big\} \times 10 \\
		0 & 0 &
		\end{pmatrix}$ \\
		\hline
	\end{tabular}
\end{center}
\caption{Training data parameters for the three simulation experiments. For each experiment, the data generating mechanism was a forward stopping ratio model.}
\label{tab:sim-settings}
\end{table}

Simulation results are shown in Table \ref{tab:sim-results}.  For a given replicate, out-of-sample prediction accuracy was evaluated as the average log-likelihood of 10,000 observations generated from the same distribution as the training set. The validation data set was chosen large enough that the within-replicate standard error of the mean was negligible relative to the across-replicate standard error. More precisely, the results were produced in the following way. For a given replicate, generate 10,000 observations from  the same distribution as the training set. Fix replicate $i$, for these 10,000 observations, compute  log-likelihoods $x_{i,j}$, $j=1,...,10000$ and denote the sample average as $v_i$,
$$
v_i  = \frac{1}{10000} \sum^{10000}_{j=1} x_{i,j}.
$$
Table \ref{tab:sim-results} reports the sample average of the $v_i$'s (across the $100$ simulated datasets), $\bar v = \frac{1}{100} \sum^{100}_{i=1} v_i$ and the standard error of $\bar v$.

For Simulation 1, the data generating model is nonparallel, and the sample size is relatively large. This results in the nonparallel and semi-parallel models having better fits than the parallel model. For Simulation 2, the data generating mechanism is parallel, and only one in three covariates has a nonzero effect. This results in the parallel and semi-parallel models having better fits than the nonparallel model. Simulation 3 is similar to Simulation 2, but the first covariate has a highly nonparallel effect. This results in the semi-parallel model having a better fit than both the parallel and nonparallel models.

Note that in every simulation scenario, the semi-parallel model fit is nearly as good, if not better, than both the parallel and nonparallel model fits. This is not to say that the semi-parallel model should always be preferred, but it is evidence that it is a highly versatile model. In practice, cross validation could be used to determine which of the three methods performs the best, using out-of-sample log-likelihood, or another measure of fit. In addition, it could be worthwhile to use cross validation to compare different values of $\rho$ for the semi-parallel fit.

\begin{table}[H]
	\begin{center}
		\include{table}
	\end{center}
	\caption{Results for the simulation experiment. For a given replicate, out-of-sample prediction accuracy was evaluated as the average log-likelihood of 10,000 observations generated from the same distribution as the training set. The values reported are the sample average (standard error) of these values across the $100$ simulation replicates.}
	\label{tab:sim-results}
\end{table}


\section{Method comparison}\label{sec:compare}
We now demonstrate ELMO class models alongside other methods for out-of-sample prediction and variable selection. We use a dataset from a cancer genomics example presented by \citet{Archer2014}. The data come from the Gene Expression Omnibus GSE18081. The CRAN package \textbf{ordinalgmifs} contains a filtered version of this dataset called \textit{hccmethyl}. It contains expression levels of 46 genes from 56 human subjects. The measurements come from liver tissue samples assayed using the Illumina GoldenGat Methylation BeadArray Cancer Panel I \citep{Archer2010}. Twenty subjects have a normal liver, 16 have cirrhosis (disease), and 20 have hepatocellular carcinoma (severe disease). These categories have a natural ordering according to disease severity.

The analysis goal was to use gene expression values to predict liver disease---more specifically, to estimate the probability that each subject's liver would be classified as healthy, diseased, or severely diseased. The number of predictors is large relative to the sample size, making regularization and/or variable selection imperative. We use five-fold cross validation to compare the out-of-sample prediction performance of various methods. We use out-of-sample log-likelihood and misclassification rate as performance criteria. A total of seven methods were compared. We summarize them below.

\begin{enumerate}
	\item Cumulative logit models with lasso penalty.
	\begin{itemize}
		\item Three models: parallel, nonparallel, and semi-parallel.
		\item $\lambda$ was tuned by five-fold cross validation within each cross validation fold, selecting the value with the best average out-of-sample log-likelihood.
		\item $\lambda$ was selected from the same sequence within each fold. $\lambda_{\max}$ was calculated from the full data, and a sequence of twenty values was generated uniformly on the logarithmic scale from $\lambda_{\max}$ to $\lambda_{\max} / 100$.
		\item Fit by \textbf{ordinalNet}.
	\end{itemize}
	\item Multinomial logistic regression models with lasso penalty.
	\begin{itemize}
		\item Two models: standard lasso and group lasso.
		\item $\lambda$ was tuned by five-fold cross validation within each cross validation fold, selecting the value with the best average out-of-sample log-likelihood.
		\item $\lambda$ was selected from the same sequence within each fold. $\lambda_{\max}$ was calculated from the full data, and a sequence of twenty values was generated uniformly on the logarithmic scale from $\lambda_{\max}$ to $\lambda_{\max} / 100$.
		\item Fit by \textbf{glmnet}.
	\end{itemize}
	\item Cumulative logit model with GMIFS solution path.
	\begin{itemize}
		\item The solution path was fit with step size $0.01$.
		\item AIC was used for tuning within each fold. Specifically, the model with the best AIC was selected from all of the models in the solution path.
		\item Fit by \textbf{ordinalgmifs}.
	\end{itemize}
	\item Cumulative logit model with AIC, forward stepwise variable selection.
	\begin{itemize}
		\item Fit by \textbf{MASS} function \texttt{polr} \citep{venables2002}.
	\end{itemize}
\end{enumerate}

Results for the experiment are summarized in Table \ref{tab:cv-results}.  The GMIFS method performed the best according to both performance criteria. The parallel cumulative logit lasso, the semi-parallel cumulative logit lasso, and the multinomial logistic regression with (standard) lasso performed comparably to the GMIFS method. The other methods did not perform well.

Poor performance of the nonparallel, cumulative logit lasso model can be attributed to the parameter space restriction discussed in Sections \ref{subsec:forms} and \ref{subsec:cp-constraint}. With this dataset, it is easy for the nonparallel model to predict non-monotone cumulative probabilities for out-of-sample observations. However, the largest $\lambda$ value in the sequence corresponds to the null model, which cannot have non-monotone cumulative probabilities. As a result, the cross validation tuning procedure tends to select the null model or a highly penalized model; neither model is good for prediction.

The AIC stepwise method performed poorly in terms of out-of-sample log-likelihood. This poor performance is  likely due to overfitting. More specifically,  this method occasionally estimates a very small probability in the observed class of an out-of-sample observation. The out-of-sample misclassification rate is more reasonable than the out-of-sample log-likelihood, but other methods perform better.

\begin{table}[H]
\begin{center}
	\include{oos-perf}
\end{center}
\caption{Out-of-sample log-likelihood and misclassification rates for the method comparison based on the liver disease dataset (GSE18081). The value reported is the mean (standard error) across five cross validation folds.}
\label{tab:cv-results}
\end{table}

In addition to class prediction, penalized regression can be used for variable selection, i.e. to determine which of the 46 genes are most associated with liver disease. Table \ref{table:coefs} shows which genes were selected by each method. All models were tuned on the full data the same way that they were trained in the cross validation study.

\begin{table}[H]
\scriptsize
\begin{center}
	\include{coefs}
\end{center}
\caption{Regression coefficient estimates for the method comparison based on the liver disease dataset (GSE18081). The column headers are abbreviations for the methods listed in the same order as Table \ref{tab:cv-results}. For the methods with multiple coefficients per predictor, the number of nonzero coefficient estimates is reported instead of the coefficient value. The nonparallel cumulative logit model can have up to two nonzero coefficients per predictor. The semi-parallel cumulative logit and multinomial logistic regression models can have up to three nonzero coefficients per predictor. By design, the group lasso penalty either selects all three coefficients or sets them all to zero.  Note that the coefficient estimates from \texttt{MASS::polr} are multiplied by -1 to be consistent with the \textbf{ordinalNet} and \textbf{ordinalgmifs} parameterizations of the cumulative logit model.}
\label{table:coefs}
\end{table}


\section{The ordinalNet R package}\label{sec:package}
The \textbf{ordinalNet} package contains the following functions:
\begin{itemize}
\item \texttt{ordinalNet} is the main function for fitting parallel, nonparallel, and semi-parallel regression models with the elastic net penalty. It returns an 'ordinalNetFit' S3 object.
\item \texttt{summary} method for 'ordinalNetFit' objects.
\item \texttt{coef} method for 'ordinalNetFit' objects.
\item \texttt{predict} method for 'ordinalNetFit' objects.
\item \texttt{ordinalNetTune} uses K-fold cross validation to obtain out-of-sample log-likelihood and misclassification rates for a sequence of lambda values.
\item \texttt{ordinalNetCV} uses K-fold cross validation to obtain out-of-sample log-likelihood and misclassification rates. Lambda is tuned within each cross validation fold.
\end{itemize}
Below is a description of the \texttt{ordinalNet} function and its arguments.

\begin{verbatim}
ordinalNet(x, y, alpha = 1, standardize = TRUE, penaltyFactors = NULL,
  positiveID = NULL, family = c("cumulative", "sratio", "cratio", "acat"),
  reverse = FALSE, link = c("logit", "probit", "cloglog", "cauchit"),
  customLink = NULL, parallelTerms = TRUE, nonparallelTerms = FALSE,
  parallelPenaltyFactor = 1, lambdaVals = NULL, nLambda = 20,
  lambdaMinRatio = 0.01, includeLambda0 = FALSE, alphaMin = 0.01,
  pMin = 1e-08, stopThresh = 1e-04, threshOut = 1e-08, threshIn = 1e-08,
  maxiterOut = 100, maxiterIn = 1000, printIter = FALSE,
  printBeta = FALSE, warn = TRUE, keepTrainingData = TRUE)
\end{verbatim}

\subsection*{Arguments}

\begin{description}

\item[x]{Covariate matrix. It is recommended that categorical covariates are
	converted to a set of indicator variables with a variable for each category
	(i.e. no baseline category); otherwise the choice of baseline category will
	affect the model fit.}
\item[y]{Response variable. Can be a factor, ordered factor, or a matrix
	where each row is a multinomial vector of counts. A weighted fit can be obtained
	using the matrix option, since the row sums are essentially observation weights.
	Non-integer matrix entries are allowed.}
\item[alpha]{The elastic net mixing parameter, with \texttt{0 <= alpha <= 1}.
\texttt{alpha=1} corresponds to the lasso penalty, and \texttt{alpha=0} corresponds
to the ridge penalty.}
\item[standardize]{If \texttt{standardize=TRUE}, the predictor variables are
scaled to have unit variance. Coefficient estimates are returned on the
original scale.}
\item[penaltyFactors]{Nonnegative vector of penalty factors for each variable.
This vector is multiplied by lambda to get the penalty for each variable.
If \texttt{NULL}, the penalty factor is one for each coefficient.}
\item[positiveID]{Logical vector indicating whether each coefficient should
be constrained to be non-negative. If \texttt{NULL}, the default value is \texttt{FALSE}
for all coefficients.}
\item[family]{Specifies the type of model family. Options are "cumulative"
for cumulative probability, "sratio" for stopping ratio, "cratio" for continuation ratio,
and "acat" for adjacent category.}
\item[reverse]{Logical. If \texttt{TRUE}, then the "backward" form of the model
is fit, i.e. the model is defined with response categories in reverse order.
For example, the reverse cumulative model with $K+1$ response categories
applies the link function to the cumulative probabilities $P(Y \ge 2),
\ldots, P(Y \ge K+1)$, rather then $P(Y \le 1), \ldots, P(Y \le K)$.}
\item[link]{Specifies the link function. The options supported are logit,
probit, complementary log-log, and cauchit. Only used if \texttt{customLink=NULL}.}
\item[customLink]{Optional list containing a vectorized link function \texttt{g},
a vectorized inverse link \texttt{h}, and the Jacobian function of the inverse link
\texttt{getQ}. The Jacobian should be defined as $\partial h(\eta) / \partial \eta^T$
(as opposed to the transpose of this matrix).}
\item[parallelTerms]{Logical. If \texttt{TRUE}, then parallel coefficient terms
will be included in the model. \texttt{parallelTerms} and \texttt{nonparallelTerms}
cannot both be \texttt{FALSE}.}
\item[nonparallelTerms]{Logical. if \texttt{TRUE}, then nonparallel coefficient terms
will be included in the model. \texttt{parallelTerms} and \texttt{nonparallelTerms}
cannot both be \texttt{FALSE}.}
\item[parallelPenaltyFactor]{Numeric value greater than or equal to zero. Lambda
is multiplied by this factor (as well as variable-specific \texttt{penaltyFactors})
to obtain the penalties for parallel terms. Only used if \texttt{parallelTerms=TRUE}.}
\item[lambdaVals]{An optional user-specified lambda sequence (vector). If \texttt{NULL},
a sequence will be generated based on \texttt{nLambda} and \texttt{lambdaMinRatio}.
In this case, the maximum lambda is the smallest value that sets all penalized
coefficients to zero, and the minimum lambda is the maximum value multiplied
by the factor \texttt{lambdaMinRatio}.}
\item[nLambda]{Positive integer. The number of lambda values in the solution path.
Only used if \texttt{lambdaVals=NULL}.}
\item[lambdaMinRatio]{A factor greater than zero and less than one. Only used
if \texttt{lambdaVals=NULL}.}
\item[includeLambda0]{Logical. If \texttt{TRUE}, then zero is added to the end
of the sequence of \texttt{lambdaVals}. This is not done by default because
it can significantly increase computational time. An unpenalized saturated model
may have infinite coefficient solutions, in which case the fitting algorithm
will still terminate when the relative change in log-likelihood becomes small.
Only used if \texttt{lambdaVals=NULL}.}
\item[alphaMin]{Value greater than zero, but much less than one.
If \texttt{alpha < alphaMin}, then \texttt{alphaMin} is used to calculate the
maximum lambda value. When \texttt{alpha=0}, the maximum lambda value would be
infinite otherwise.}
\item[pMin]{Value greater than zero, but much less than one. During the optimization
routine, the Fisher information is calculated using fitted probabilities. For
this calculation, fitted probabilities are capped below by this value to prevent
numerical instability.}
\item[stopThresh]{In the relative log-likelihood change between successive
lambda values falls below this threshold, then the last model fit is used for all
remaining lambda.}
\item[threshOut]{Convergence threshold for the coordinate descent outer loop.
The optimization routine terminates when the relative change in the
penalized log-likelihood between successive iterations falls below this threshold.
It is recommended to set \texttt{theshOut} equal to \texttt{threshIn}.}
\item[threshIn]{Convergence threshold for the coordinate descent inner loop. Each
iteration consists of a single loop through each coefficient. The inner
loop terminates when the relative change in the penalized approximate
log-likelihood between successive iterations falls below this threshold.
It is recommended to set \texttt{theshOut} equal to \texttt{threshIn}.}
\item[maxiterOut]{Maximum number of outer loop iterations.}
\item[maxiterIn]{Maximum number of inner loop iterations.}
\item[printIter]{Logical. If \texttt{TRUE}, the optimization routine progress is
printed to the terminal.}
\item[printBeta]{Logical. If \texttt{TRUE}, coefficient estimates are printed
after each coordinate descent outer loop iteration.}
\item[warn]{Logical. If \texttt{TRUE}, the following warning message is displayed
when fitting a cumulative probability model with \texttt{nonparallelTerms=TRUE}
(i.e. nonparallel or semi-parallel model).
``Warning message: For out-of-sample data, the cumulative probability model
with nonparallelTerms=TRUE may predict cumulative probabilities that are not
monotone increasing.''
The warning is displayed by default, but the user may wish to disable it.}
\item[keepTrainingData]{Logical. If \texttt{TRUE}, then \texttt{x} and \texttt{y}
are saved with the returned "ordinalNetFit" object. This allows
\texttt{predict.ordinalNetFit} to return fitted values for the training data
without passing a \texttt{newx} argument.}
\end{description}


\section{Demonstration in R}\label{sec:demo}
This section contains six examples that demonstrate different aspects of the \textbf{ordinalNet} package. Specifically, using the GSE18081 dataset from the Gene Expression Omnibus, we demonstrate how the penalized ELMO class models can be used for prediction and variable selection.

\subsection*{Example 1}
We illustrate how to fit the Gene Expression Omnibus GSE18081 data using \textbf{ordinalNet}. We fit a cumulative probability model with logit link (proportional odds model). We use the default settings \texttt{parallelTerms=TRUE} and \texttt{nonparallelTerms=FALSE} to fit the parallel model. We use the default elastic net tuning parameter \texttt{alpha=1} to select the lasso penalty. We use the default settings of \texttt{lambdaVals=NULL}, \texttt{nLambda=20}, and \texttt{lambdaMinRatio=1e-2} to generate a sequence of twenty $\lambda$ values, with $\lambda_{\text{max}}$ equal to the smallest value that sets every coefficient to zero and $\lambda_{\text{min}}=\lambda_{\text{max}} \cdot 0.01$. The sequence runs from $\lambda_{\text{min}}$ to $\lambda_{\text{max}}$ uniformly on the logarithmic scale.

The \texttt{summary} method displays the lambda sequence (\texttt{lambdaVals}), number of nonzero coefficients (\texttt{nNonzero}), the log-likelihood (\texttt{loglik}), percent deviance explained (\texttt{pctDev}), and AIC and BIC. The AIC and BIC are calculated using \texttt{nNonzero} as the approximate degrees of freedom. The \texttt{coef} method returns the coefficient estimates of any model fit in the sequence---the best AIC fit is selected by default. The \texttt{matrix=TRUE} option returns the coefficients in matrix form with a column corresponding to each linear predictor. Because \texttt{fit1} is a parallel model, the coefficient columns are identical except for the intercepts.

\begin{verbatim}
R> library(ordinalNet)
R> library(ordinalgmifs)  # contains hccmethyl data
R> library("Biobase")  # contains functions pData and exprs
R> data(hccmethyl)
R> y <- pData(hccmethyl)$group
R> x <- t(exprs(hccmethyl))
R> 
R> fit1 <- ordinalNet(x, y, family="cumulative", link="logit")
R> 
R> head(summary(fit1))

  lambdaVals nNonzero    loglik    devPct       aic      bic
1  0.4287829        2 -61.22898 0.0000000 126.45797 130.5087
2  0.3364916        6 -49.70793 0.1881634 111.41586 123.5680
3  0.2640652       10 -40.97485 0.3307932 101.94970 122.2032
4  0.2072278       11 -33.86289 0.4469467  89.72579 112.0047
5  0.1626241       12 -28.29049 0.5379560  80.58097 104.8852
6  0.1276209       15 -23.15157 0.6218855  76.30313 106.6834

R> head(coef(fit1, matrix=TRUE))

                     logit(P[Y<=1]) logit(P[Y<=2])
(Intercept)              -27.997567     -19.157113
CDKN2B_seq_50_S294_F     -13.774058     -13.774058
DDIT3_P1313_R             -8.393522      -8.393522
ERN1_P809_R                1.215556       1.215556
GML_E144_F                 7.263032       7.263032
HDAC9_P137_R               0.000000       0.000000
\end{verbatim}

\subsection*{Example 2}
By setting \texttt{parallelTerms=TRUE} and \texttt{nonparallelTerms=TRUE}, we obtain the semi-parallel model fit for \texttt{fit2}. Because this is a cumulative probability model, we set \texttt{warn=FALSE} to suppress the warning that the semi-parallel form is susceptible to non-monotone cumulative probabilities for out-of-sample predictions. We use the default semi-parallel tuning parameter $\rho$, which is \texttt{parallelPenaltyFactor=1}. The coefficient matrix of the best AIC fit has nearly identical columns for the first several predictors, but they differ for the first predictor.

\begin{verbatim}
R> fit2 <- ordinalNet(x, y, family="cumulative", link="logit",
R+                    parallelTerms=TRUE, nonparallelTerms=TRUE, warn=FALSE)
R> 
R> head(coef(fit2, matrix=TRUE))

                     logit(P[Y<=1]) logit(P[Y<=2])
(Intercept)              -23.518682     -22.199966
CDKN2B_seq_50_S294_F      -5.732730     -18.218945
DDIT3_P1313_R             -8.604492      -8.604492
ERN1_P809_R                1.010048       1.010048
GML_E144_F                 7.414796       7.414796
HDAC9_P137_R               0.000000       0.000000
\end{verbatim}

\subsection*{Example 3}
We now demonstrate the problem with the nonparallel cumulative probability model discussed in Sections \ref{subsec:forms} and \ref{subsec:cp-constraint}. (The semi-parallel model is also susceptible to this issue, but it is less prone. It does not occur for the semi-parallel model on this data set). As seen from the \texttt{summary} method output, the solution path is terminated after the third $\lambda$ value where the optimum leaves the constrained parameter space. Better solutions must exist for the remaining $\lambda$ values, but the coordinate descent optimization procedure is not designed handle this issue.

\begin{verbatim}
R> fit3 <- ordinalNet(x, y, family="cumulative", link="logit",
R+                    parallelTerms=FALSE, nonparallelTerms=TRUE, warn=FALSE)
R> 
R> head(summary(fit3))

  lambdaVals nNonzero    loglik    devPct      aic      bic
1  0.4046054        2 -61.22898 0.0000000 126.4580 130.5087
2  0.3175182        4 -52.35095 0.1449972 112.7019 120.8033
3  0.2491755        6 -44.89072 0.2668388 101.7814 113.9335
4  0.1955430        6 -44.89072 0.2668388 101.7814 113.9335
5  0.1534543        6 -44.89072 0.2668388 101.7814 113.9335
6  0.1204248        6 -44.89072 0.2668388 101.7814 113.9335
\end{verbatim}

\subsection*{Example 4}
The \texttt{ordinalNetTune} function uses $K$-fold cross validation to obtain out-of-sample performance for a sequence on $\lambda$ values. We use the default setting of \texttt{nFolds=5} and the default sequence of twenty $\lambda$ values obtained from the model fit to the full data. The user can use this information to tune the model, for example by selecting the $\lambda$ value with the best average out-of-sample likelihood across folds, as demonstrated below.

\begin{verbatim}
R> set.seed(123)
R> fit2_tune <- ordinalNetTune(x, y, family="cumulative", link="logit",
R+                             parallelTerms=TRUE, nonparallelTerms=TRUE, 
R+                             warn=FALSE, printProgress=FALSE)
R> 
R> head(fit2_tune$loglik)

             fold1      fold2      fold3      fold4      fold5
lambda1 -13.131358 -11.810575 -12.031408 -11.701774 -11.906341
lambda2 -10.642481  -9.965017 -10.385299 -10.553901 -11.220306
lambda3  -8.743227  -8.532953  -9.132148  -9.240154  -9.815121
lambda4  -7.529487  -7.027589  -8.157234  -8.185017  -8.756163
lambda5  -6.982274  -5.883983  -7.172058  -7.307804  -7.733298
lambda6  -6.494204  -4.880251  -6.102234  -6.544728  -6.851243

R> bestLambdaIndex <- which.max(rowMeans(fit2_tune$loglik))
R> head(coef(fit2_tune$fit, matrix=TRUE, whichLambda=bestLambdaIndex))

                     logit(P[Y<=1]) logit(P[Y<=2])
(Intercept)              -15.998499     -14.993664
CDKN2B_seq_50_S294_F      -8.526472     -12.769316
DDIT3_P1313_R             -6.132272      -6.132272
ERN1_P809_R                1.660266       1.660266
GML_E144_F                 5.153204       5.153204
HDAC9_P137_R               0.000000       0.000000
\end{verbatim}

\subsection*{Example 5}
The \texttt{ordinalNetCV} function uses $K$-fold cross validation to obtain out-of-sample performance of models that are tuned within each cross validation fold. We use the default value of \texttt{nFolds=5} and the same default $\lambda$ sequence as above. We also use the default settings of \texttt{nFoldsCV=5} and \texttt{tuneMethod="cvLoglik"} to tune $\lambda$ by 5-fold cross validation within each fold, each time selecting the $\lambda$ value with the best average out-of-sample log-likelihood.

We compare the performance of the parallel, semi-parallel, and nonparallel cumulative probability models. By default, the \texttt{ordinalNetCV} function will randomly create the fold partitions, but we pass a list of fold partitions to ensure that the same partitions are used for each of the three methods. (This could also be done by setting the same seed before each of the three calls to \texttt{ordinalNetCV}). Although not of critical importance, we also increase the outer iteration limit to 200 for the semi-parallel model.

We see that the parallel and semi-parallel models have similar performance, but the nonparallel model is much worse. This is because the nonparallel model out-of-sample log-likelihood becomes undefined (non-monotone cumulative probabilities) within the first few $\lambda$ values. Further examination of the $\lambda$ index value selected for each fold reveals that often the first $\lambda$ value (i.e. the null model) is selected for the nonparallel model. It likely does not matter that the solution path is terminated for leaving the parameter space because the cross validation procedure would probably select very large $\lambda$ values even if the entire solution path were available.

\begin{verbatim}
R> set.seed(123)
R> nFolds <- 5
R> n <- nrow(x)
R> indexRandomOrder <- sample(n)
R> folds <- split(indexRandomOrder, rep(1:nFolds, length.out=n))
R> 
R> fit1_cv <- ordinalNetCV(x, y, family="cumulative", link="logit",
R+                         parallelTerms=TRUE, nonparallelTerms=FALSE,
R+                         printProgress=FALSE, folds=folds)
R> 
R> fit2_cv <- ordinalNetCV(x, y, family="cumulative", link="logit",
R+                         parallelTerms=TRUE, nonparallelTerms=TRUE, warn=FALSE,
R+                         printProgress=FALSE, folds=folds, maxiterOut=200)
R> 
R> fit3_cv <- ordinalNetCV(x, y, family="cumulative", link="logit",
R+                         parallelTerms=FALSE, nonparallelTerms=TRUE, warn=FALSE,
R+                         printProgress=FALSE, folds=folds)
R> 
R> loglik123 <- cbind(fit1_cv$loglik, fit2_cv$loglik, fit3_cv$loglik)
R> bestLambdaIndex123 <- cbind(fit1_cv$bestLambdaIndex, fit2_cv$bestLambdaIndex, 
R+                             fit3_cv$bestLambdaIndex)
R> colnames(loglik123) <- colnames(bestLambdaIndex123) <- c("fit1_cv", "fit2_cv", 
R+                                                          "fit3_cv")
R> loglik123

         fit1_cv   fit2_cv   fit3_cv
fold1 -3.7368745 -3.418520 -13.06444
fold2 -0.6340903 -1.123142 -11.67827
fold3 -0.9521387 -1.267511  -8.65993
fold4 -2.6007935 -2.231969 -11.99773
fold5 -3.7018103 -2.899580 -11.15106

R> colMeans(loglik123)

   fit1_cv    fit2_cv    fit3_cv 
 -2.325141  -2.188144 -11.310287 

R> bestLambdaIndex123

      fit1_cv fit2_cv fit3_cv
fold1      17      19       1
fold2      19      13       1
fold3      18      15       3
fold4      13      14       1
fold5      11      13       2
\end{verbatim}

\subsection*{Example 6}
Finally, we use the same cross validation procedure to evaluate the performance of the reverse stopping ratio family. This family does not have parameter constraints, which is an issue unique to the cumulative probability family. In this case, the nonparallel model is much more competitive, but the parallel and semi-parallel models still have a slight edge. It appears that deviation from the parallelism assumption is not too substantial in this population. Perhaps with a larger sample size, the flexibility of the semi-parallel and nonparallel models would be more beneficial.

\begin{verbatim}
R> set.seed(123)
R> 
R> fit4_cv <- ordinalNetCV(x, y, family="sratio", link="logit", reverse=TRUE,
R+                         parallelTerms=TRUE, nonparallelTerms=FALSE,
R+                         printProgress=FALSE, folds=folds)
R> 
R> fit5_cv <- ordinalNetCV(x, y, family="sratio", link="logit", reverse=TRUE,
R+                         parallelTerms=TRUE, nonparallelTerms=TRUE,
R+                         printProgress=FALSE, folds=folds)
R> 
R> fit6_cv <- ordinalNetCV(x, y, family="sratio", link="logit", reverse=TRUE,
R+                         parallelTerms=FALSE, nonparallelTerms=TRUE,
R+                         printProgress=FALSE, folds=folds)
R> 
R> loglik456 <- cbind(fit4_cv=fit4_cv$loglik, fit5_cv=fit5_cv$loglik, 
R+                    fit6_cv=fit6_cv$loglik)
R> colnames(loglik456) <- c("fit4_cv", "fit5_cv", "fit6_cv")
R> loglik456

         fit4_cv    fit5_cv   fit6_cv
fold1 -3.6642141 -3.4503556 -3.038360
fold2 -0.8693706 -0.7925935 -1.530216
fold3 -1.2445013 -1.1095928 -1.791319
fold4 -1.1022247 -3.2404628 -3.519975
fold5 -2.9746301 -2.4929934 -2.175200

R> colMeans(loglik456)

  fit4_cv   fit5_cv   fit6_cv 
-1.970988 -2.217200 -2.411014
\end{verbatim}

\section{Discussion}\label{sec:discussion}
This paper introduced the elmentwise link, multinomial-ordinal (ELMO) model class, a rich class of multinomial regression models that includes commonly used categorical regression models. Each of these models has both a parallel and nonparallel form. The parallel form is appropriate for ordinal data, while the nonparallel form is a more flexible model which can be used with an unordered categorical response. We also introduced  the semi-parallel model form, which can be used with the elastic net penalty to shrink the nonparallel model toward the parallel model.

The motivation for this work began with a need to extend variable selection tools for ordinal logistic regression. For instance, consider the problem of developing a gene signature to predict response to a novel therapy, where the observed patient response belongs to one of the following categories: no response, partial response, or complete response. We developed these tools in the general ELMO framework.  Specifically, we proposed a coordinate descent fitting algorithm for the ELMO class with the elastic net penalty. The algorithm is general and can also be applied to multinomial regression models outside the ELMO class.

We considered numerical experiments to highlight different features of the model class and to demonstrate the use of the related \textsf{R} code. We presented different simulation scenarios to demonstrate cases where the parallel, nonparallel, and semi-parallel each achieved better out-of-sample prediction performance than the other two models. With the Gene Expression Omnibus GSE18081 data set, we demonstrated the use of the penalized ELMO class  for prediction and variable selection.

Finally, we introduced the \textsf{R} package \textbf{ordinalNet}, which implements a coordinate descent algorithm for parallel, nonparallel, and semi-parallel models of the ELMO class. It is available on the Comprehensive \textsf{R} Archive Network at \url{http://CRAN.R-project.org/package=ordinalNet}.

We consider two possible directions for future research: code speedup via \CC  and questions of statistical inference. \textbf{Rcpp} and \textbf{RcppArmadillo} are \textsf{R} packages which allow integration of \CC  code into \textsf{R} \citep{eddelbuettel2011rcpp, eddelbuettel2013seamless, eddelbuettel2014rcpparmadillo}. Our code is written with separate  functions for the inner and outer coordinate descent loops. Because of the number of calls to it in a typical run of the algorithm, the inner loop, in particular, could benefit from speed up via \CC.

The \textbf{ordinalNet} package does not provide standard errors for estimates. We quote a relevant section from the \textbf{penalized} vignette \citep{goeman2014}.
\begin{quote}
It is a very natural question to ask for standard errors of regression coefficients or other estimated quantities. In principle such standard errors can easily be calculated, e.g. using the bootstrap. Still, this package deliberately does not provide them. The reason for this is that standard errors are not very meaningful for strongly biased estimates such as arise from penalized estimation methods. Penalized estimation is a procedure that reduces the variance of estimators by introducing substantial bias. The bias of each estimator is therefore a major component of its mean squared error, whereas its variance may contribute only a small part.
\end{quote}
The topic of post-selection inference has been studied in both the classic setting \citep{zhang1992inference,leeb2005model,wang2009inference,berk2013valid}, where the number of observations exceeds the number of predictors, and the high-dimensional setting \citep{javanmard2014confidence, lockhart2014significance,tibshirani2016exact}. In the high-dimensional setting, we would like to highlight the groundbreaking work of \citet{lockhart2014significance}. They proved the asymptotic distribution of their test statistic specifically for the linear model, but their simulation results suggest that the same test statistic could be used for generalized linear models. This work may provide a path for post-selection inference for penalized multinomial and ordinal regression models.

\section*{Acknowledgements}
The authors thank Alex Tahk for the suggestion that led them to explore shrinking the nonparallel model to the parallel model. This work was supported in part by NIH grants T32HL083806 and HG007377.

\nocite{RCoreTeam2017}
\bibliographystyle{ims}
\bibliography{ordinal}

\appendix

\section{Uniqueness of the semi-parallel model estimator}\label{app:uniqueness}
This section addresses the uniqueness problem for the semi-parallel model. The semi-parallel model without the elastic net penalty is not identifiable, as there are infinitely many \ps for any particular model. However, different \ps have different elastic net penalty terms; therefore, the penalized likelihood favors some \ps over others. We will demonstrate that amongst almost all \ps for a given model, the elastic net penalty has a unique optimum; hence, the penalized likelihood has a unique optimum. There is one exception: the lasso penalty with integer-valued $\rho$. In this case, there may be a small range of optima on a closed interval.

We proceed in the following manner. First, we formulate the basic problem. Next we consider uniqueness for the ridge penalty ($\alpha = 0$), which is the simplest case.  We then consider the lasso penalty ($\alpha = 1$), where this exception can occur. Finally, we consider the elastic net penalty with $\alpha \in (0,1)$. These derivations are related to derivations for the lasso in the linear model setting \citep{osborne2000lasso, rinaldo,tibshirani2013lasso}.

To formulate the problem, take any row of the semi-parallel model coefficient matrix $(B_{j1}, B_{j2}, \ldots, B_{jK})$ and the corresponding component of the coefficient vector, $b_j$. Denote their values as $(\delta_1, \delta_2, \ldots, \delta_K)$ and $\zeta$, respectively. For any set of values $(\beta_1, \beta_2, \ldots, \beta_K)$, there are an infinite number of \ps such that
$$
(\delta_1 + \zeta, \delta_2 + \zeta, \ldots, \delta_K + \zeta) = (\beta_1, \beta_2, \ldots, \beta_K).
$$
To see this, for any $\zeta$  set $\delta_k = \beta_k - \zeta$ for all $k$. All of these \ps have the same likelihood because they specify the same model, but they have different elastic net penalty terms proportional to
$$
\alpha \left(\rho |\zeta| + \sum_{k=1}^K |\delta_k|\right) + \tfrac{1}{2} (1-\alpha) \left(\rho \zeta^2 + \sum_{k=1}^K \delta_k^2\right).
$$
Our goal is to find the value of $\zeta$ that minimizes the elastic net penalty.

We solve this as a constrained optimization problem, minimizing the penalty over $(\zeta, \delta_1, \delta_2, \ldots, \delta_K)$ subject to the constraints $\delta_1 + \zeta = \beta_1$, $\delta_2 + \zeta = \beta_2$, \ldots, $\delta_K + \zeta = \beta_K$. This is equivalent to minimizing the Lagrangian
\begin{equation*}
\begin{split}
L(\zeta, \delta_1, \delta_2, \ldots, \delta_K, \lambda_1, \lambda_2, \ldots, \lambda_K) &= \alpha \left(\rho |\zeta| + \sum_{k=1}^K |\delta_k|\right) + \tfrac{1}{2} (1-\alpha) \left(\rho \zeta^2 + \sum_{k=1}^K \delta_k^2\right) \\
&+ \lambda_1 (\delta_1 + \zeta - \beta_1) + \lambda_2 (\delta_2 + \zeta - \beta_1) + \cdots + \lambda_K (\delta_K + \zeta - \beta_K) \ .
\end{split}
\end{equation*}

\subsubsection*{Ridge regression}
In this case, the Lagrangian  is differentiable everywhere. Consider
$$
0 \overset{\text{set}}{=} \frac{\partial L}{\partial \delta_k} = \delta_k + \lambda_k = \beta_k - \zeta + \lambda_k.
$$
Solving this yields
$$
\lambda_k = \zeta - \beta_k.
$$
Now consider,
$$
0 \overset{\text{set}}{=} \frac{\partial L}{\partial \zeta} = \rho \zeta + \lambda_1 + \lambda_2 + \cdots + \lambda_K = \rho \zeta + K \zeta - (\beta_1 + \beta_2 + \cdots + \beta_K).
$$
Solving this yields
$$
\zeta = \frac{1}{K + \rho} (\beta_1 + \beta_2 + \cdots + \beta_K).
$$


The solution is unique for any $\rho \ge 0$. \\

\subsubsection*{Lasso}
Consider
$$
0 \overset{\text{set}}{=} \frac{\partial L}{\partial \delta_k} = \text{sign}(\delta_k) + \lambda_k = I\{\beta_k > \zeta\} - I\{\beta_k < \zeta\} +\lambda_k.
$$
Solving this yields
$$
\lambda_k = I\{\beta_k < \zeta\} - I\{\beta_k > \zeta\}.
$$
Next, consider
\begin{equation*}
\begin{split}
\frac{\partial L}{\partial \zeta} &= \rho \cdot \text{sign}(\zeta) + \lambda_1 + \lambda_2 + \cdots + \lambda_K \\
&= \rho \cdot \text{sign}(\zeta) - \left( \sum I\{\beta_k > \zeta\} - \sum I\{\beta_k < \zeta\} \right).
\end{split}
\end{equation*}


We want the solution where $\frac{\partial L}{\partial \zeta}$ equals or crosses zero. That is, we are searching for the value of $\zeta$ where $f(\zeta)$ equals or crosses $\rho$, with
$$
f(\zeta ) = \text{sign}(\zeta) \cdot \left( \sum I\{\beta_k > \zeta\} - \sum I\{\beta_k < \zeta\} \right).
$$
Note that if $\rho \ge K$, then the solution will be $\zeta = 0$. Hence, if $\rho \ge K$, then all parallel coefficients will be penalized to zero, and the fit will be equivalent to the nonparallel model with the elastic net penalty.

Now, if $\rho$ is not an integer, then the solution is unique. If $\rho$ is an integer, then the solution could be unique, or there may be a range of solutions on a closed interval between two consecutive $\beta$'s, ranked by value.

\subsubsection*{Elastic net}
Consider
$$
0 \overset{\text{set}}{=} \frac{\partial L}{\partial \delta_k} = \alpha \cdot \text{sign}(\delta_k) + (1-\alpha) \delta_k + \lambda_k = \alpha \cdot \left( I\{\beta_k > \zeta\} - I\{\beta_k < \zeta\} \right) + (1-\alpha) (\beta_k - \zeta) + \lambda_k .
$$
Solving this yields
$$
\lambda_k = \alpha \cdot \left( I\{\beta_k < \zeta\} - I\{\beta_k > \zeta\} \right) + (1-\alpha) (\zeta - \beta_k) .
$$
Now, consider
\begin{equation*}
\begin{split}
\frac{\partial L}{\partial \zeta} &=  \rho \alpha \cdot \text{sign}(\zeta) + \rho (1-\alpha) \zeta + \lambda_1 + \lambda_2 + \cdots + \lambda_K \\
&= \rho \alpha \cdot \text{sign}(\zeta) + \rho (1-\alpha) \zeta - \alpha \left( \sum I\{\beta_k > \zeta\} - \sum I\{\beta_k < \zeta\} \right) - \\
&\quad -(1-\alpha) (\beta_1 + \beta_2 + \cdots + \beta_K - K \zeta) \\
&= \rho \alpha \cdot \text{sign}(\zeta) + (1-\alpha) (\rho + K) \zeta - \alpha \left( \sum I\{\beta_k > \zeta\} - \sum I\{\beta_k < \zeta\} \right) - \\
&\quad - (1-\alpha) (\beta_1 + \beta_2 + \cdots + \beta_K - K \zeta).
\end{split}
\end{equation*}

We want the solution where $\frac{\partial L}{\partial \zeta}$ equals or crosses zero. This solution is less transparent than that of ridge or lasso. However, the partial derivative is piecewise linear in $\zeta$ and never constant over a range of values. Hence, the solution is unique.


\section{The inverse link function and its Jacobian for specific MO families}\label{app:mo}
Each MO family is defined by the $g_{MO}(p)$ component of its multivariate link function. For optimization, it is necessary to compute the inverse $h_{MO}(\delta)$ and its Jacobian $D h_{MO}(\delta)$. In this section, we provide a method to compute these three functions for the cumulative probability, stopping ratio, continuation ratio, and adjacent category families. In some cases, it is convenient to compute the elements of these functions recursively. Although the Jacobian is, strictly speaking, a function of $\delta$, we write it in terms of both $p$ and $\delta$ when convenient. This can be done because there is a one-to-one correspondence between $\delta$ and $p$.

Each family has a forward and backward form. We present only one of these forms for each family. To fit the backward form, one can simply define the response categories in reverse order and fit the forward model, and vice versa.

\subsection*{Forward cumulative probability family}
This family is defined by $\delta_j = P(Y \le j)$ ( see Table \ref{tab:mo}). From this definition, for all $j$,
$$
[g_{MO}(p)]_j = \sum_{i=1}^j p_i.
$$
Now, $h_{MO}$ has a closed form with
$$
[h_{MO}(\delta)]_j = \delta_j - \delta_{j-1}, \quad \text{for all $j$}.
$$
$D h_{MO}$ also has a closed form with
\begin{equation*}
[D h_{MO}(\delta)]_{mn} =
	\begin{cases}
	\delta_m (1-\delta_m) & m=n \\
	-\delta_m (1-\delta_m) & n = m-1 \\
	0 & \text{otherwise}.
	\end{cases} \
\end{equation*}

\subsection*{Forward stopping ratio family}
This family is defined by $\delta_j = P(Y = j | Y \ge j)$ (see Table \ref{tab:mo}).  From this definition, $[g_{MO}(p)]_1 = p_1$ and, for $j=2, \ldots, K$,
\begin{equation*}
[g_{MO}(p)]_j = \frac{p_j }{1 -\sum_{i=1}^{j-1} p_i} \ .
\end{equation*}
$h_{MO}(\delta)$ can be computed recursively, beginning with $[h_{MO}(\delta)]_1 = \delta_1$. For $j=2, \ldots, K$,
\begin{equation*}
[h_{MO}(\delta)]_j = \delta_j \left( 1 - \sum_{i=1}^{j-1} [h_{MO}(\delta)]_i \right) \ .
\end{equation*}

$D h_{MO}(\delta)$ can also be computed recursively. For the first row we have
$$
[D h_{MO}(\delta)]_{1 \cdot} = (1, 0, \ldots, 0).
$$
For $m=2, \ldots K$,
\begin{equation*}
[D h_{MO}(\delta)]_{m \cdot} = -\delta_m \sum_{i=1}^{m-1} [Dh_{MO}(\delta)]_{i \cdot} + \left( 1-\sum_{i=1}^{m-1} [h_{MO}(\delta)]_i \right) \cdot (0, \ldots, 0, \underset{m^\text{th}}{1}, 0, \ldots, 0) \ .
\end{equation*}

\subsection*{Forward continuation ratio family}
This family is defined by $\delta_j = P(Y > j | Y \ge j)$ (Table \ref{tab:mo}). Let $g_{MO:FSR}$, $h_{MO:FSR}$, and $D h_{MO:FSR}$ denote link, inverse link and inverse link Jacobian, respectively, for the forward stopping ratio family. Using these function definitions, it is straightforward to compute the corresponding functions for the forward continuation ratio family. We have
\begin{eqnarray*}
g_{MO}(p) &=& \mathbbm{1} - g_{MO:FSR}(p),\\
h_{MO}(\delta) &=& h_{MO:FSR}(\mathbbm{1} - \delta),\\
D h_{MO}(\delta) &=& -D h_{MO:FSR}(\mathbbm{1} - \delta).
\end{eqnarray*}


\subsection*{Forward adjacent category family}
This family is defined by $\delta_j = P(Y = j+1 | j \le Y \le j+1)$ (see Table \ref{tab:mo}). From this definition, for all $j$, we have
$$
[g_{MO}(p)]_j = \frac{p_{j+1}}{p_j + p_{j+1}}.
$$

Now, let $\Delta_j = \delta_j / (1-\delta_j)$. $h_{MO}(\delta)$ can be computed recursively, beginning with
$$
[h_{MO}(\delta)]_1 = \dfrac{1}{1 + \sum_{i=1}^K \prod_{j=1}^i \Delta_j} .
$$
For $j = 2, \ldots, K$,
\begin{equation*}
[h_{MO}(\delta)]_j = [h_{MO}(\delta)]_{j-1} \Delta_{j-1}.
\end{equation*}

To compute $D h_{MO}(\delta)$, we write $p = (p_1, p_1 \Delta_1, p_2 \Delta_2, \ldots, p_{K-1} \Delta_{K-1})$. $D h_{MO}(\delta)$ can also be computed recursively. Beginning with the first row, we have
\begin{equation*}
[D h_{MO}(\delta)]_{1 \cdot} = -\frac{p_1 (1-p_1)}{\Delta_1}, -\frac{p_1 (1-p_1-p_2)}{\Delta_2}, \ldots,  -\frac{p_1 (1-p_1-\cdots-p_K)}{\Delta_K}.
\end{equation*}
Then, for $m = 2, \ldots, K$,
\begin{equation*}
[D h_{MO}(\delta)]_{m \cdot} = \Delta_{m-1} [D h_{MO}(\delta)]_{(m-1) \cdot} + p_{m-1} (0, \ldots, 0, \underset{m^\text{th}}{1}, 0, \ldots, 0) .
\end{equation*}


\section{Quadratic approximation to the log-likelihood}\label{sec:quadapp}
The quadratic approximation $\ell^{(r)}(\beta)$ is the second order Taylor expansion of $\ell$ at $\hat{\beta}^{(r)}$, up to an additive constant that does not depend on $\beta$. Also, in  $\ell^{(r)}(\beta)$, the Hessian is replaced by its expectation, $-\mathcal{I}(\hat{\beta}^{(r)})$. We give the derivation below, where $\overset{\text{C}}{=}$ denotes equality up to an additive constant. We have
\begin{equation*}
	\begin{split}
		&\ell(\hat{\beta}^{(r)}) + (\beta - \hat{\beta}^{(r)})^T U(\hat{\beta}^{(r)}) - \tfrac{1}{2} (\beta - \hat{\beta}^{(r)})^T \ \mathcal{I}(\hat{\beta}^{(r)}) \ (\beta - \hat{\beta}^{(r)}) \\
	&\overset{\text{C}}{=} \beta^T U(\hat{\beta}^{(r)}) + \beta^T \mathcal{I}(\hat{\beta}^{(r)}) \hat{\beta}^{(r)} - \tfrac{1}{2} \beta^T \mathcal{I}(\hat{\beta}^{(r)}) \beta \\
		&\overset{\text{C}}{=} \beta^T X^T W^{(r)} ([W^{(r)}]^{-1} v^{(r)} + X \hat{\beta}^{(r)}) - \tfrac{1}{2} \beta^T X^T W^{(r)} X \beta \\
		&\overset{\text{C}}{=} \beta^T X^T W^{(r)} z^{(r)} - \tfrac{1}{2} \beta^T X^T W^{(r)} X \beta - \tfrac{1}{2} [z^{(r)}]^T W^{(r)} z^{(r)} \\
		&= -\tfrac{1}{2} (z^{(r)} - X \beta)^T\ W^{(r)}\ (z^{(r)} - X \beta) \\
		&= \ell^{(r)}(\beta).
	\end{split}
\end{equation*}


\section{Statistical inference}\label{sec:inference}
\citet{goeman2014} assert that while standard errors of penalized regression coefficients could be calculated, e.g. by the bootstrap, these standard errors should be interpreted with care. In particular, they should not be used to construct confidence intervals. This is because penalized regression coefficient estimators have substantial bias. For this reason, they deliberately do not provide standard error estimates in the \textbf{penalized} package. In the classical setting, where the number of observations exceeds the number of predictors, then one can use the usual likelihood-based methods to construct confidence intervals and hypothesis tests. However, even in this setting, such methods can lead to incorrect inference \citep{zhang1992inference, leeb2005model,wang2009inference,berk2013valid}.

\citet{Tibshirani2011}, \citet{lockhart2014significance}, and \citet{Tibshirani2016} laid the groundwork of an asymptotic inference method for penalized regression. They proved the asymptotic distribution of their test statistic specifically for least squares regression, but their simulation results suggest that the same test statistic could be used for generalized linear models. This may be a path toward high dimensional inference for penalized multinomial and ordinal regression models.

%% file: table.tex
\begin{tabular}{rlll}
  \hline
 & Parallel & Nonparallel & Semi-parallel \\ 
  \hline
Sim 1 & -1.05 (0.00045) & -0.95 (0.00052) & -0.95 (0.00052) \\ 
  Sim 2 & -0.59 (0.0089) & -0.71 (0.0073) & -0.62 (0.0077) \\ 
  Sim 3 & -0.74 (0.008) & -0.71 (0.010) & -0.64 (0.011) \\ 
   \hline
\end{tabular}

%% file: oos-perf.tex
\begin{tabular}{lrr}
  \hline
 & Log-likelihood & Misclassification rate \\ 
  \hline
Cumulative logit lasso - parallel & -2.37 (0.60) & 0.091 (0.050) \\ 
  Cumulative logit lasso - nonparallel & -10.92 (0.70) & 0.373 (0.036) \\ 
  Cumulative logit lasso - semi-parallel & -2.47 (0.52) & 0.091 (0.050) \\ 
  Multinomial logistic lasso & -2.84 (0.41) & 0.108 (0.019) \\ 
  Multinomial logistic group lasso & -6.56 (0.49) & 0.158 (0.047) \\ 
  Cumulative logit GMIFS & -1.76 (0.58) & 0.073 (0.034) \\ 
  Cumulative logit AIC forward stepwise & -12.87 (4.03) & 0.162 (0.035) \\ 
   \hline
\end{tabular}

%% file: coefs.tex
\begin{tabular}{rrrrrrrr}
  \hline
 & CLP & CLN & CLS & ML & MLG & GMIFS & AIC \\ 
  \hline
CDKN2B\_seq\_50\_S294\_F & -14.74 & $\cdot$ & 2 & 1 & 3 & -1.30 & $\cdot$ \\ 
  DDIT3\_P1313\_R & -8.88 & $\cdot$ & 1 & $\cdot$ & $\cdot$ & -1.29 & -148.94 \\ 
  ERN1\_P809\_R & 1.20 & $\cdot$ & 1 & $\cdot$ & $\cdot$ & 0.36 & $\cdot$ \\ 
  GML\_E144\_F & 7.84 & $\cdot$ & 1 & 1 & 3 & 1.92 & $\cdot$ \\ 
  HDAC9\_P137\_R & $\cdot$ & $\cdot$ & $\cdot$ & $\cdot$ & $\cdot$ & 0.08 & $\cdot$ \\ 
  HLA-DPA1\_P205\_R & $\cdot$ & $\cdot$ & $\cdot$ & $\cdot$ & $\cdot$ & 0.36 & $\cdot$ \\ 
  HOXB2\_P488\_R & $\cdot$ & $\cdot$ & $\cdot$ & $\cdot$ & 3 & -0.08 & -171.45 \\ 
  IL16\_P226\_F & 14.43 & $\cdot$ & 1 & 1 & 3 & 1.82 & $\cdot$ \\ 
  IL16\_P93\_R & 2.25 & $\cdot$ & 1 & $\cdot$ & $\cdot$ & 0.34 & 77.05 \\ 
  IL8\_P83\_F & 1.36 & $\cdot$ & 2 & 1 & 3 & 0.38 & $\cdot$ \\ 
  MPO\_E302\_R & 11.88 & $\cdot$ & 1 & 1 & 3 & 0.72 & 188.97 \\ 
  MPO\_P883\_R & $\cdot$ & $\cdot$ & 1 & 1 & $\cdot$ & 0.17 & $\cdot$ \\ 
  PADI4\_P1158\_R & -4.40 & $\cdot$ & 1 & 1 & 3 & -0.96 & $\cdot$ \\ 
  SOX17\_P287\_R & -9.41 & $\cdot$ & 1 & 2 & 3 & -1.94 & $\cdot$ \\ 
  TJP2\_P518\_F & -24.62 & $\cdot$ & 1 & 1 & 3 & -2.05 & -71.70 \\ 
  WRN\_E57\_F & 5.75 & $\cdot$ & 1 & $\cdot$ & $\cdot$ & 0.54 & 145.49 \\ 
  CRIP1\_P874\_R & $\cdot$ & 1 & $\cdot$ & 1 & 3 & $\cdot$ & $\cdot$ \\ 
  SLC22A3\_P634\_F & $\cdot$ & 1 & 1 & 2 & 3 & $\cdot$ & $\cdot$ \\ 
  CCNA1\_P216\_F & $\cdot$ & $\cdot$ & $\cdot$ & $\cdot$ & $\cdot$ & $\cdot$ & $\cdot$ \\ 
  SEPT9\_P374\_F & $\cdot$ & $\cdot$ & $\cdot$ & $\cdot$ & $\cdot$ & $\cdot$ & $\cdot$ \\ 
  ITGA2\_E120\_F & $\cdot$ & $\cdot$ & $\cdot$ & $\cdot$ & $\cdot$ & $\cdot$ & $\cdot$ \\ 
  ITGA6\_P718\_R & $\cdot$ & $\cdot$ & $\cdot$ & $\cdot$ & $\cdot$ & $\cdot$ & $\cdot$ \\ 
  HGF\_P1293\_R & $\cdot$ & $\cdot$ & $\cdot$ & $\cdot$ & $\cdot$ & $\cdot$ & $\cdot$ \\ 
  DLG3\_E340\_F & $\cdot$ & $\cdot$ & $\cdot$ & $\cdot$ & $\cdot$ & $\cdot$ & $\cdot$ \\ 
  APP\_E8\_F & $\cdot$ & $\cdot$ & $\cdot$ & $\cdot$ & $\cdot$ & $\cdot$ & $\cdot$ \\ 
  SFTPB\_P689\_R & 5.05 & $\cdot$ & 1 & 1 & 3 & 0.31 & $\cdot$ \\ 
  PENK\_P447\_R & $\cdot$ & $\cdot$ & $\cdot$ & $\cdot$ & $\cdot$ & $\cdot$ & $\cdot$ \\ 
  COMT\_E401\_F & 3.74 & $\cdot$ & 1 & $\cdot$ & 3 & 0.59 & $\cdot$ \\ 
  NOTCH1\_E452\_R & $\cdot$ & $\cdot$ & $\cdot$ & $\cdot$ & $\cdot$ & $\cdot$ & $\cdot$ \\ 
  EPHA8\_P456\_R & $\cdot$ & $\cdot$ & $\cdot$ & $\cdot$ & $\cdot$ & $\cdot$ & $\cdot$ \\ 
  WT1\_P853\_F & $\cdot$ & $\cdot$ & $\cdot$ & $\cdot$ & $\cdot$ & $\cdot$ & $\cdot$ \\ 
  KLK10\_P268\_R & $\cdot$ & $\cdot$ & $\cdot$ & 1 & 3 & $\cdot$ & $\cdot$ \\ 
  PCDH1\_P264\_F & 0.08 & $\cdot$ & 1 & 1 & 3 & $\cdot$ & $\cdot$ \\ 
  TDGF1\_P428\_R & $\cdot$ & $\cdot$ & $\cdot$ & $\cdot$ & $\cdot$ & $\cdot$ & $\cdot$ \\ 
  EFNB3\_P442\_R & $\cdot$ & $\cdot$ & $\cdot$ & $\cdot$ & $\cdot$ & $\cdot$ & -137.05 \\ 
  MMP19\_P306\_F & $\cdot$ & $\cdot$ & $\cdot$ & $\cdot$ & $\cdot$ & $\cdot$ & $\cdot$ \\ 
  FGFR2\_P460\_R & $\cdot$ & $\cdot$ & $\cdot$ & 1 & 3 & $\cdot$ & $\cdot$ \\ 
  RAF1\_P330\_F & $\cdot$ & $\cdot$ & $\cdot$ & $\cdot$ & $\cdot$ & $\cdot$ & $\cdot$ \\ 
  BMPR2\_E435\_F & $\cdot$ & $\cdot$ & $\cdot$ & 1 & 3 & $\cdot$ & $\cdot$ \\ 
  GRB10\_P496\_R & $\cdot$ & $\cdot$ & $\cdot$ & $\cdot$ & $\cdot$ & $\cdot$ & $\cdot$ \\ 
  CTSH\_P238\_F & $\cdot$ & $\cdot$ & $\cdot$ & $\cdot$ & $\cdot$ & $\cdot$ & $\cdot$ \\ 
  SLC6A8\_seq\_28\_S227\_F & $\cdot$ & $\cdot$ & $\cdot$ & $\cdot$ & $\cdot$ & $\cdot$ & $\cdot$ \\ 
  PLXDC1\_P236\_F & $\cdot$ & $\cdot$ & $\cdot$ & $\cdot$ & $\cdot$ & $\cdot$ & $\cdot$ \\ 
  TFE3\_P421\_F & $\cdot$ & $\cdot$ & $\cdot$ & $\cdot$ & $\cdot$ & $\cdot$ & $\cdot$ \\ 
  TSG101\_P139\_R & $\cdot$ & $\cdot$ & $\cdot$ & $\cdot$ & $\cdot$ & $\cdot$ & $\cdot$ \\ 
   \hline
\end{tabular}